 \documentclass[showpacs,preprintnumbers,amsmath,amssymb,nofootinbib]{revtex4}
\usepackage{epsfig}
\newcommand \be {\begin{equation}}
\newcommand \ee {\end{equation}}
\newcommand \bea {\begin{eqnarray}}
\newcommand \eea {\end{eqnarray}}
\def\roughly#1{\mathrel{\raise.3ex\hbox
{$#1$\kern-.75em\lower1ex\hbox{$\sim$}}}}
\def\lsim{\roughly<}
\def\gsim{\roughly>}

\begin{document}
\title{Modified Friedmann Equation and Inflation in Warped 
Codimension-two Braneworld}
\author{Fang Chen}
\author{James M. Cline}
\author{Sugumi Kanno}
\affiliation{Physics Department, McGill University\\
3600 University Street, Montreal, QC, Canada H3A 2T8}
\date{\today}
\begin{abstract}

We study the Friedmann equation for the warped codimension-two
braneworld background which most closely resembles the
Randall-Sundrum model.  Extra matter on the (Planck) 4-brane, with
equation of state $p_\theta=(\alpha-1)\rho$ for the azimuthal
pressure, is required to satisfy the junction conditions.  For $1 <
\alpha < 5$, we show that there are two static solutions to the
Einstein equations for given values of the brane stress-energies.  
Close to the static solutions, the relation between Hubble expansion
rate $H$ and brane tension reproduces the standard 4D result for
small $H$, but exhibits unusual deviations when $H$ is of order the
AdS curvature scale.  The two static branches for $1 <\alpha < 5$ are
shown to come together smoothly at a maximum value of $H$; however
the radion is shown to be unstable in the branch with higher $H$.  
This remains true even with a mechanism for stabilization of the
radion, {\it i.e.,} the Goldberger-Wise (GW) mechanism, since large
enough $H$ overcomes the force of stabilization. Even in the
unstabilized case, cosmological constraints on the time and spatial
variation of Newton's constant are typically satisfied; only fifth
force constraints require the stabilization.  For $\alpha > 5$ the
model is intrinsically stable, without the need for a GW field, and
in this case we show that inflationary predictions can be modified by
the nonstandard Friedmann equation; in particular it is possible to
get an upper limit on the spectral index, large deviations from the
consistency condition between the tensor spectrum and ratio $r$, and
large running of the spectral index even though the slow roll
parameters remain small. 

\end{abstract}

\maketitle
\section{Introduction}

The cosmology of braneworld scenarios has been widely studied,
particularly in the simplest case of codimension-one branes,
involving only a single extra dimension 
(see \cite{Langlois,Maartens,Brax,Kanno}
for reviews). It was initially noticed
that the Friedmann equation was strongly modified from its usual 
dependence $H^2\sim \rho$ to the form $H\sim\rho$ \cite{BDL}, in
contradiction to big bang nucleosynthesis and other cosmological
tests.  Before the realization that such pathological behavior was
linked to the failure to stabilize the extra dimension, 
it was discovered that in the
Randall-Sundrum (RS) model \cite{RSI,RSII} with a warped extra 
dimension due to a
negative bulk cosmological constant, the Friedmann equation took a
more interesting form,
\begin{eqnarray}
	H^2=\frac{8\pi G}{3}\rho\left(1+\frac{\rho}{T}\right),
\label{mfe}
\end{eqnarray}
where $T$ is the brane tension~\cite{CGKT,Cline,Binetruy}.  
Thus the exotic linear dependence
$H\sim\rho$ could arise as a high-energy correction to the normal
low-energy dependence.  The behavior (\ref{mfe}) is only valid for 
the single-brane version of RS; in the case of two branes,
it is necessary to stabilize the interbrane separation \cite{CGRT},
which leads to a more complicated dependence for the high-energy
corrections \cite{CV}.  Other interesting forms are possible for the
Friedmann equation in codimension-one brane models, as
in the DGP model~\cite{Dvali}, whose Lagrangian includes an extra Einstein-Hilbert
term localized on the brane, and leads to a quadratic equation
determining $H$,
\begin{eqnarray}
	H^2=\frac{8\pi G}{3}\rho\mp\frac{2M_5^3}{M_4^2}H,
\label{5dfei}
\end{eqnarray}
in terms of the 5D Planck mass $M_5$ and 4D Planck mass $M_4$. More intricate possibilities
also exist, for example through the addition of a 
Gauss-Bonnet term to the DGP model \cite{Brown:2005ug}.

There has also been considerable interest in 6D, codimension-two
braneworld models, in part motivated by the
suggestion of a self-tuning mechanism for the cosmological constant
which was argued to be a special feature of codimension-two branes
\cite{Carroll,Burgess:2001bn,Leblond:2001xr,CDGV}, 
although this claim is controversial
\cite{Vinet,Papa}.  Regardless of the cosmological constant problem,
it is still interesting to consider the predictions of
codimension-two branes for the expansion of the early universe, since
deviations from the general-relativistic prediction can help
constrain the models, as well as lead to new possibilities for 
inflation \cite{Maartens:1999hf,CLL}.

In the present work we revisit the problem of modifications to the
Friedmann equation, focusing on a warped model which is the natural
extension of the RS model to six spacetime dimensions 
\cite{Send}-\cite{regularized}.  
Part of our motivation is a claim \cite{CDGV} that the standard
Friedmann equation is not recovered even when the extra dimensions
are stabilized.  We will show that this claim was erroneous, and 
that although the unstabilized version of the model indeed exhibits
unusual expansion behavior at high energies, the expected results of
general relativity (GR) are recovered at low energy in the case where
the radion is stabilized through a Goldberger-Wise (GW) \cite{GW}
mechanism, and even in the {\it unstabilized} model.  

The detailed form of the modified Friedmann equation depends on the
equation of state $\alpha$ of extra matter which is placed on the 4-brane
that plays the role of the UV brane in our model.  We find that for
most values of $\alpha$, there is an exotic branch of the Friedmann
equation which has the Hubble rate $H$ increasing as the brane
tension {\it decreases}.  Through a fluctuation analysis, we show
that the radion is unstable on this branch, even in the version of 
the model which includes the GW mechanism.  

On the conventional branch of the Friedmann relation, we obtain
interesting deviations from the predictions of GR near (but still
below) the point where the radion instability begins.  We show that
it is possible to obtain potentially observable deviations from the 
consistency condition which relates the tensor spectral index to the
tensor-to-scalar ratio, in a model of chaotic inflation on the brane.
We note that there is a suppression of the spectral index in the
braneworld inflation model relative to standard chaotic inflation. 
Furthermore it is  possible (though it requires more fine tuning) to
get large running of the spectral index.

The organization of this paper is as follows. In section \ref{model},
the basic setup is presented. We derive the bulk equation of  motion
and the jump conditions for the codimension two background with the
3-brane undergoing de Sitter expansion. We obtain an exact solution
in this background. In section \ref{Feq}, we study the Friedmann
equation in the model in the absence of radion stabilization.  The
Friedmann equation is obtained using an analytical
approximation valid at low $H$, as well as numerically for arbitrary
$H$. We find that standard cosmological expansion is recovered at low
$H$, but interesting deviations occur when $H\sim \ell$, where $\ell$
is the curvature scale of the AdS background. In section \ref{FEGW}
we repeat the analysis in the more realistic version of the model
which includes stabilization of the bulk using the Goldberger-Wise
mechanism.  To check the effectiveness of the stabilization as a
function of the Hubble rate,  we analyze the stability of the model,
with and without the GW field, in section \ref{STB}. To explore the
physical consequences of the modified Friedmann equation, we study
chaotic inflation on the brane in section \ref{IFL}. Section
\ref{CON} gives our conclusions, including a summary of the most
important results of the paper. In appendix \ref{DD}, details of
the derivation of the perturbative approximation to the modified Friedmann equation
are supplied.  Appendix \ref{sec:symbols} gives an index of the 
many symbols used in this paper as an aid to the bewildered.

\section{The Model}
\label{model}

We will consider a codimension-two braneworld with a negative
cosmological constant in the six-dimensional bulk spacetime.   The
geometry describes a warped conical throat, which is bounded at large
$r$,  namely $r=P$, by a 4-brane around which  orbifold boundary
conditions are imposed, analogous to the Planck brane of the RS
model.  In addition there is a 3-brane at the infrared end of the
throat, at the position $r=\varrho$, which can  be thought of as the
Standard Model brane.  The action is
\begin{eqnarray}
S&=&\frac{1}{2k_6^2}\int d^{\,6}x\sqrt{-{\cal G}}
\left[\,{\cal R}-2\Lambda_6\,\right]
-\int d^{\,6}x\sqrt{-{\cal G}}\left[\,
\frac{1}{2}\,(\nabla\phi)^2+V(\phi)\,\right]
\cr
\cr
&&+\int d^{\,5}x\sqrt{-\tilde{g}}\,{\cal L}_{4\hbox{-}{\rm brane}}
-\int d^{\,4}x\sqrt{-g}\,\tau_3\,,
\label{theaction}
\end{eqnarray}
where $k_6^2$ is the 6-dimensional gravitational constant and 
${\cal G}_{AB}$ is the 6-dimensional bulk metric.
The 6D cosmological constant is negative,
$\Lambda_6=-10/\ell^2$, and gives rise to an approximately AdS$_6$
bulk geometry with curvature length scale $\ell$.  
 We denote the
induced metrics on the 3-brane and the 4-brane by $g_{\mu\nu}$ and
$\tilde{g}_{ab}$, respectively. $\tau_3$ is the 3-brane tension.
Here, ${\cal L}_{4\hbox{-}{\rm brane}}$ is the Lagrangian
density of the matter on the 4-brane that includes
the 4-brane tension $T_4$, as well as some additional component which
is necessary for satisfying Israel matching (jump) conditions at
$r=P$.  The
4-brane at $r=P$ is needed in order to have a compact bulk and
localized gravity.   

The 6-dimensional Einstein equation derived by varying the above action
with respect to ${\cal G}^{AB}$ takes the form
\begin{eqnarray}
\sqrt{-{\cal G}}\left[G_{AB}+\Lambda_6{\cal G}_{AB}
-k_6^2T_{AB}\right]
=k_6^2\sqrt{-\tilde{g}}\,S_{ab}\,\delta^a_A\,\delta^b_B\,
\delta(r-P)
-k_6^2\tau_3\sqrt{-g}\,g_{\mu\nu}\,\delta^\mu_A\,\delta^\nu_B\,
\delta^{(2)}(r-\varrho)\,,
\label{einstein}
\end{eqnarray}
where $\delta^{(2)}$ denotes the 2-dimensional delta function
with support at the position of the 3-brane,
$S_{ab}$ is the 4-brane stress-energy tensor (which we will
specify below, eq.~(\ref{stresstensor})) and $T_{AB}$ is the 
bulk stress-energy tensor,
\begin{eqnarray}
T_{AB}=\partial_A\phi\partial_B\phi
-\frac{1}{2}{\cal G}_{AB}\partial^D\phi\partial_D\phi
-V(\phi){\cal G}_{AB}\,.
\end{eqnarray}

The pure gravity
model has a modulus, the radion, which must be stabilized to make a
realistic model \cite{Burgess:2001bn}.  For this reason we have included a bulk scalar
field, which can give the radion a mass by the Goldberger-Wise
mechanism \cite{GW}, provided that $\phi$ couples to the 4-brane
in ${\cal L}_{4\hbox{-}{\rm brane}}$.

\subsection{Bulk equations of motion}
\label{bem}

For simplicity we impose azimuthal symmetry on the extra two
dimensions and therefore require that the metric depends only on the
radial coordinate. The line element has the form
\begin{eqnarray}
ds^2=a(r)\,[\,-dt^2+e^{2\tilde Ht}\delta_{ij}dx^idx^j\,]
+f(r)K^2\,d\theta^2
+\frac{1}{f(r)}dr^2,
\label{metric1}
\end{eqnarray}
where $\tilde H$ is a rescaled Hubble parameter,\footnote{Since $a(\varrho)\neq 1$ at the 3-brane, $t$ is not
the proper time.  The Hubble rate with respect to the proper time
is $H= \tilde H/\sqrt{a(\varrho)}$, hence our distinction between $H$ and
$\tilde H$.}
 and $K$ is a dimensionful parameter which determines  the deficit angle
at the 3-brane, where there is generically a conical singularity.  To
insure that the position $r=\varrho$ of the 3-brane represents a single
point in the extra dimensions, we require that $f(\varrho)=0$.   The pure gravity
model has a modulus, the radion, which must be stabilized to make a
realistic model.  For this reason we have included a bulk scalar
field, which can give the radion a mass by the Goldberger-Wise
mechanism \cite{GW}, provided that $\phi$ couples to the 4-brane
in ${\cal L}_{\rm 4-brane}$. 

The Einstein equations for the model are
\begin{eqnarray}
&&\mu\mu:\qquad 
\frac{3\tilde{H}^2}{af}-\frac{3}{2}\frac{a^{\prime\prime}}{a}
-\frac{1}{2}\frac{f^{\prime\prime}}{f}-\frac{3}{2}\frac{a^\prime f^\prime}{af}
=\frac{\Lambda_6}{f}+k_6^2\left(\frac{1}{2}\phi^{\prime^2}+\frac{V(\phi)}{f}
\right)\nonumber\\
&&\theta\theta:\qquad
\frac{6\tilde{H}^2}{af}-2\frac{a^{\prime\prime}}{a}
-\frac{1}{2}\left(\frac{a^\prime}{a}\right)^2-\frac{a^\prime f^\prime}{af}
=\frac{\Lambda_6}{f}+k_6^2\left(\frac{1}{2}\phi^{\prime^2}+\frac{V(\phi)}{f}
\right)\nonumber\\
&&rr:\qquad
\frac{6\tilde{H}^2}{af}-\frac{a^\prime f^\prime}{af}
-\frac{3}{2}\left(\frac{a^\prime}{a}\right)^2
=\frac{\Lambda_6}{f}-k_6^2\left(\frac{1}{2}\phi^{\prime^2}
-\frac{V(\phi)}{f}\right)
\label{BEQ}
\end{eqnarray}
where $\prime$ denotes the derivative with respect to $r$.

The variation with respect to $\phi$ gives the Klein-Gordon equation,
\begin{eqnarray}
\square\,\phi-\frac{dV}{d\phi}=0\,.
\end{eqnarray}

\subsection{Jump conditions}
\label{jumpcond}
To completely specify the geometry, we must consider the jump
conditions which relate the metric and bulk scalar to the sources
of stress-energy at the boundaries.  
First, let us consider the jump condition for the 3-brane.
Recalling that the position of 3-brane satisfies
\begin{eqnarray}
	f(\varrho)=0 \label{3-brane} \ ,
\end{eqnarray}
the deficit angle is given by
\begin{eqnarray}
	\Delta\theta = 
	2\pi\left[1-\frac{K}{2}f^\prime(r)\right]\Bigg|_{r=\varrho}
\end{eqnarray}
A nonvanishing deficit angle indicates the presence of
a conical singularity at the position of the 3-brane, related to
the brane tension $\tau_3$ by $\Delta\theta = k_6^2\tau_3$.
The jump condition at $r=\varrho$ can thus be written as
\begin{eqnarray}
K=\frac{2}{f'(\varrho)}\left(1-k_6^2\frac{\tau_3}{2\pi}\right).
\label{JC:3-brane}
\end{eqnarray}
In what follows, it will not be necessary to include any coupling
of the scalar field to the 3-brane, so the boundary condition
of $\phi$ at $r=\varrho$ is simply
\be
	\phi'(\varrho) = 0
\ee	

Next, for the 4-brane, we define the stress-energy tensor of the
4-brane $S_{ab}$ as:
\begin{eqnarray}
S_{\mu\nu}&=&-\left(T_4+\frac{\tau_4}{L^\alpha_4}\right)
\tilde{g}_{\mu\nu}
\ ,\qquad
L_4^2=fK^2\nonumber\\
S_{\theta\theta}&=&-\left(T_4+\left(1-\alpha\right)
\frac{\tau_4}{L^{\alpha}_{4}}\right)\tilde{g}_{\theta\theta}\,.
\label{stresstensor}
\end{eqnarray}
The parameter $\alpha$, which allows for a difference between the
4D and the $\theta\theta$ components of the 4-brane stress tensor,
must be nonzero to accommodate a static solution with $P<\infty$.
The physics which could account for the $\tau_4$ contribution
to the stress tensor can  
arise in several ways \cite{Leblond:2001xr,CN}.   Conservation
of $S_{ab}$ implies that the energy density
of the source $\tau_4$ must scale like
$1/L_4^\alpha$ \cite{CDGV,Burgess:2001bn}
where $2\pi L_4$ is the circumference of the 4-brane.
If $\tau_4$ is due to the Casimir effect of massless fields living
on the 4-brane, then  $\alpha=5$,
whereas 
if it is due to smearing a 3-brane around the 4-brane then
$\alpha=1$.

We assume $Z_2$ orbifold boundary conditions at the 4-brane,
so that the radial derivatives of the metric components change
sign as one crosses the 4-brane, and their discontinuity is determined
by the 4-brane stress energy components.
Thus the jump conditions at $r=P$ are given by
\begin{eqnarray}
&&f\left[\,
\frac{3}{2}\left(\frac{a^\prime}{a}\right)^\prime
+\frac{1}{2}\left(\frac{f^\prime}{f}\right)^\prime
\,\right]\tilde{g}_{\mu\nu}
=k_6^2\sqrt{f}\,S_{\mu\nu}\,\delta(r-P)\,,\\
&&2\,f^2K^2\left(\frac{a^\prime}{a}\right)^\prime
=k_6^2\sqrt{f}\,S_{\theta\theta}\,\delta(r-P)\,,
\end{eqnarray}
where the term $\sqrt{f}$ in the r.h.s.\ of each equation comes
from the ratio of the determinant of the 4-brane induced metric to
the bulk metric, $\sqrt{-\tilde{g}}/\sqrt{-{\cal G}}$.

By integrating the above equations across $r=P$, assuming
$Z_2$ orbifold symmetry, we get
\begin{eqnarray}
&&\sqrt{f}\left(3\frac{a^\prime}{a}+\frac{f^\prime}{f}\right)
=k_6^2\left(T_4+\frac{\tau_4}{L_4^{\alpha}}\right)\,,
\label{JC:4-brane:1}\nonumber\\
&&4\sqrt{f}\,\frac{a^\prime}{a}
=k_6^2\left(T_4+(1-\alpha)\frac{\tau_4}{L_4^{\alpha}}\right)\,,
\label{JC:4-brane:2}
\end{eqnarray}

To stabilize the model with the bulk scalar field, we will allow for
the possibility that $\phi$ couples to the 4-brane through a potential
$V_P(\phi)$ contained in ${\cal L}_{\rm 4-brane}$.  This leads
to the boundary condition   
\be
	\phi^{\prime}=-\frac{1}{2f(P)}\frac{dV_P}{d\phi}
\ee
at $r=P$, using the assumed $Z_2$ orbifold symmetry. 

\subsection{Background solution}
The general solution to Einstein's equations for the metric
(\ref{metric1}) is
\begin{eqnarray}
a(r)=\frac{r^2}{r_0^2},\qquad\,
f(r)=\left(\frac{r^2}{r_0^2}-\frac{r_1^3}{r^3}+\tilde H^2\ell^2 \right)
\frac{r_0^2}{\ell^2}
\label{metric10}
\end{eqnarray}
The constants of integration $r_0$, $r_1$ can be set to convenient
values by rescaling coordinates,
$x^\mu \to A x^\mu, \ r\to B r  $.
Under this tranformation, the metric functions change as
\be
a\to aA^2 B^2,\qquad \tilde H\to \tilde H/A,\qquad f\to f/B^2,
\qquad K\to KB
\label{xform}
\ee
By choosing $A=(\frac{r_0}{r_1})^{3/5}$ and $B=(r_0^2r_1^3)^{1/5}$,
we can thus set $r_0=r_1=\ell$, for convenience.  To simplify the 
equations however, we will express $r$ in units of $\ell$ henceforth,
so that $r$ (as well as $\varrho, P$) becomes dimensionless. In the limit of 
$\tilde{H}\rightarrow 0$, the metric (\ref{metric10}) reduces to
 the AdS soliton solution~\cite{Horowitz:1998ha}.

Since we have chosen to normalize $a(\varrho) = \varrho^2$ at the 3-brane
instead of the usual value, $a=1$ the observed Hubble rate on the 
3-brane is given by
$H=\tilde H/\varrho$, and the position of  the 3-brane, defined by
eq.~(\ref{3-brane}), satisfies
\begin{eqnarray}
\label{rhoeq}
1+{H}^2\ell^2=\frac{1}{\varrho^5} \label{3-brane:2}\ ,
\end{eqnarray}
The jump condition for the 3-brane, eq.~(\ref{JC:3-brane}), is then
\begin{eqnarray}
K=\frac{\ell}{\varrho+\frac{3}{2\varrho^4}}\left(1-k_6^2\frac{\tau_3}{2\pi}\right)
\label{JC2:3-brane}\ ,
\end{eqnarray}
and the jump conditions for the 4-brane,
eqs.~(\ref{JC:4-brane:1}), can be expressed as
\begin{eqnarray}
&&F\equiv\frac{P+\frac32 P^{-4}}{\sqrt{P^2-P^{-3}+H^2\varrho^2\ell^2}}
+\frac{3\alpha+1}{\alpha-1}\frac{\sqrt{P^2-P^{-3}+H^2\varrho^2\ell^2}}{P}
=\frac{k_6^2\alpha}{2(\alpha-1)}\ell T_4
\label{JC2:4-brane:1}\\
&&G\equiv\frac{P+\frac32 P^{-4}}{\sqrt{P^2-P^{-3}+H^2\varrho^2\ell^2}}-
\frac{\sqrt{P^2-P^{-3}+H^2\varrho^2\ell^2}}{P}=
\frac{\ell k_6^2\alpha \tau_4}{2K^{\alpha}(P^2-P^{-3}+H^2\varrho^2\ell^2)^{\alpha/2}}
\label{JC2:4-brane:2}
\end{eqnarray}
These conditions cannot be solved analytically for general values
of the Hubble rate.  However, we can solve them when $H=0$,
as we next demonstrate.

\subsection{Static solutions}
\label{staticsol}

\begin{figure}[htp]\vspace{1cm}
\centerline{\includegraphics[height=5cm]{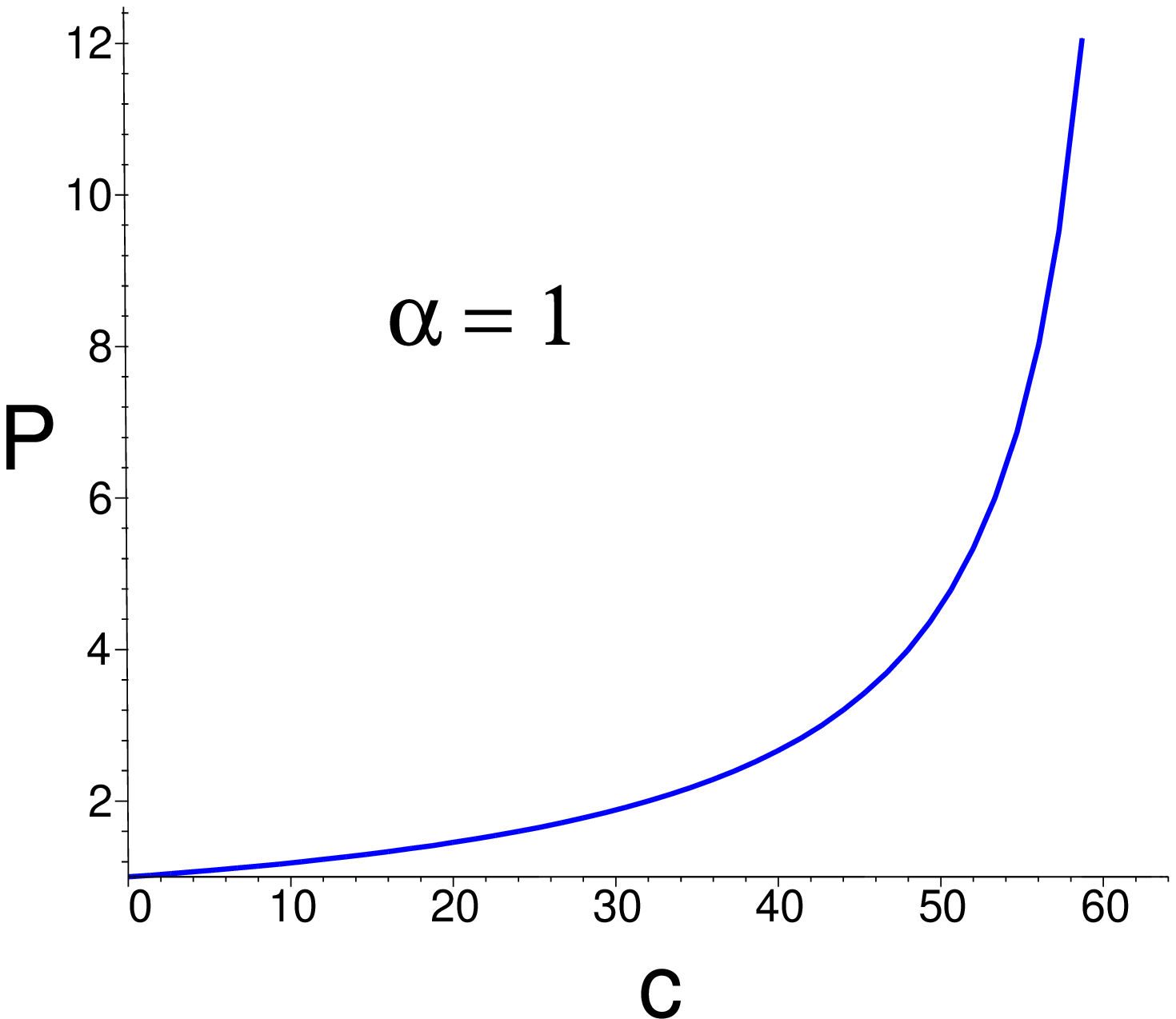}
\includegraphics[height=5cm]{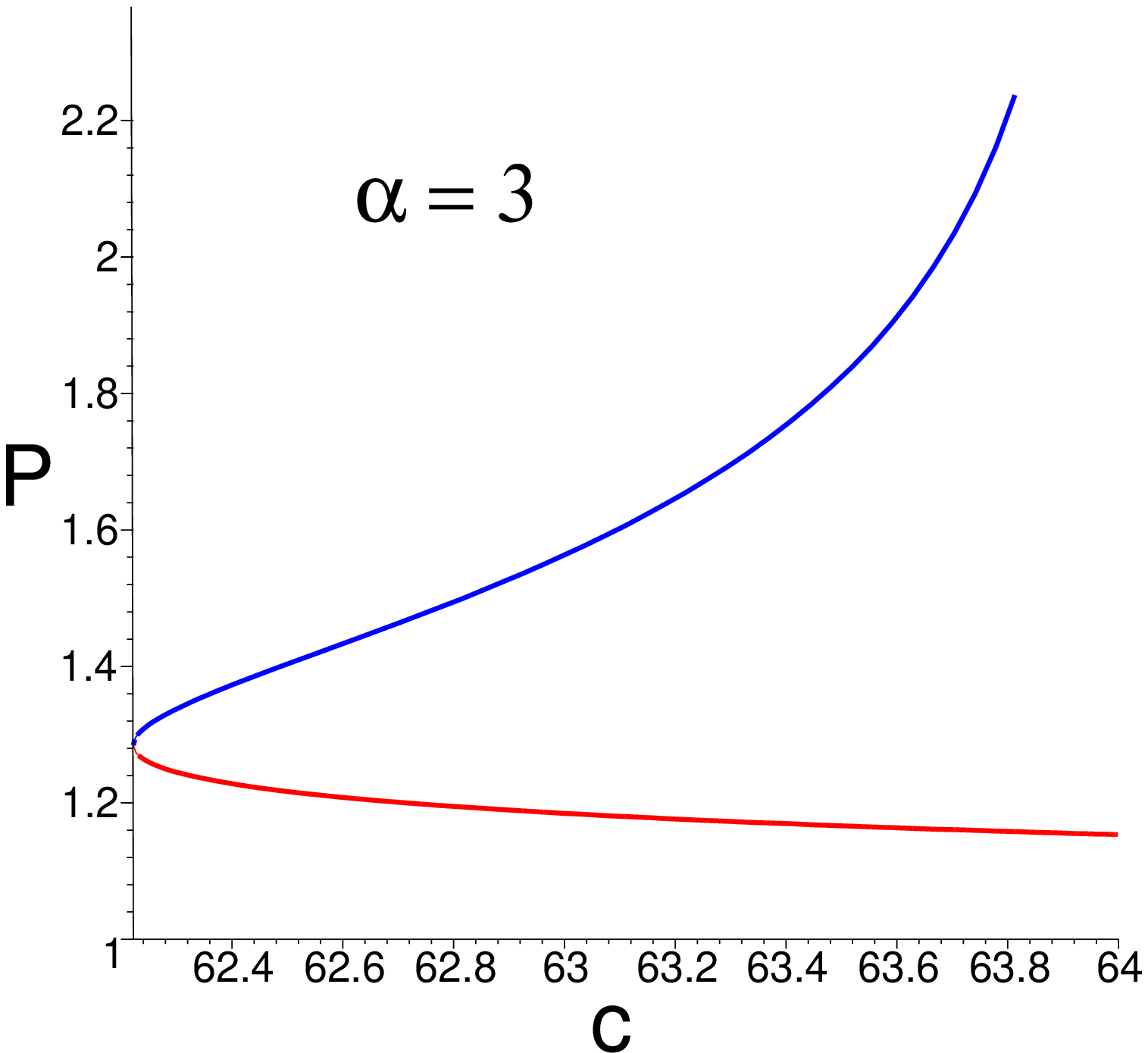}
\includegraphics[height=5cm]{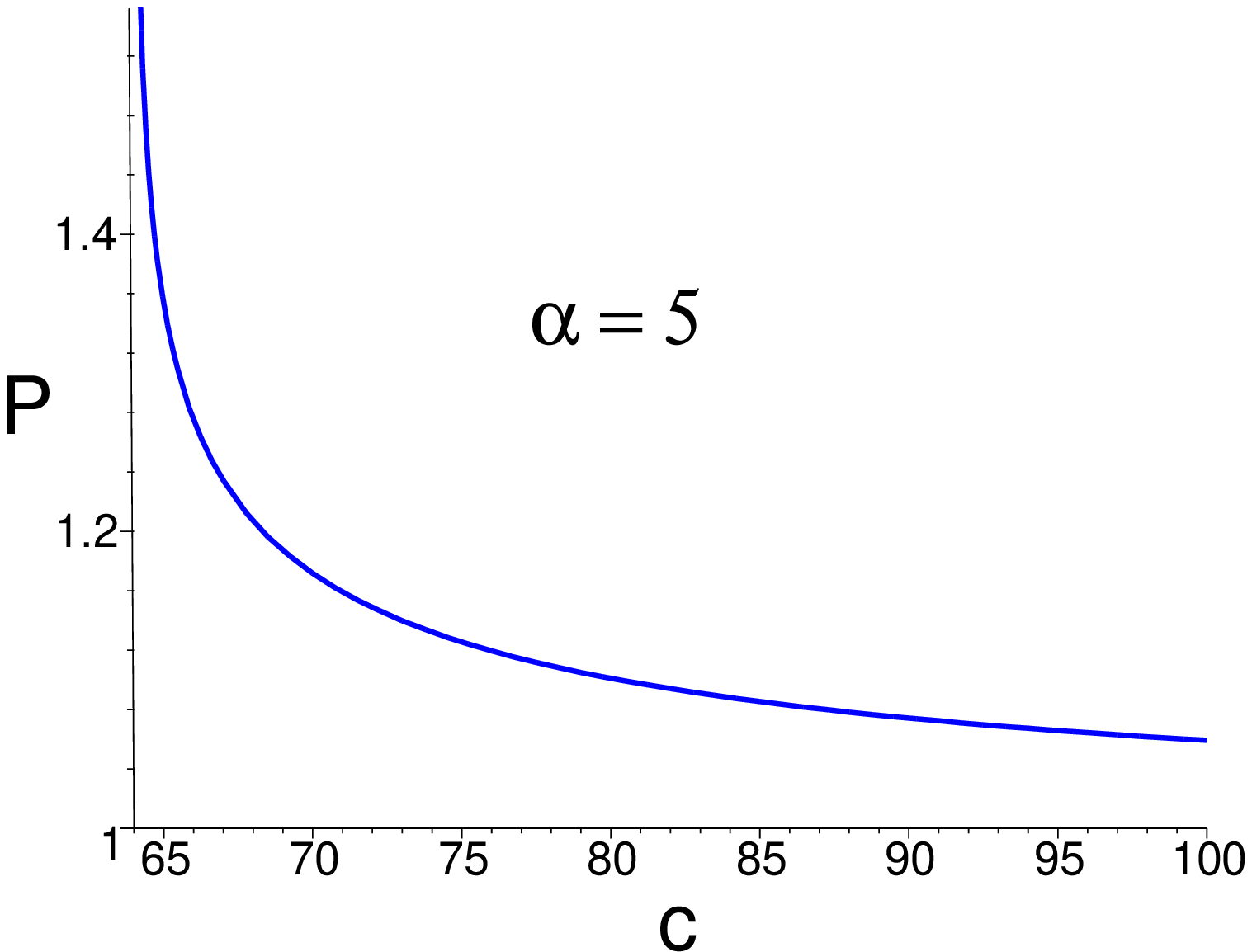}}
\caption{The solutions for $\bar P$ as a function of
 $c\equiv (k_6^2\ell T_4)^2$ for the
cases of $\alpha=1,3,5$, where $\alpha$ determines equation of state of extra
matter on the 4-brane.  For $1 < \alpha < 5$ there are two solutions
for $\bar P$, as exemplified by the middle graph.}
\label{pbar}
\end{figure}

We wish to find the value of
the 4-brane position $\bar{P}$ and the critical 3-brane tension 
$\bar\tau_3$
which satisfy the jump conditions for
a static geometry, with $H=0$: 
\begin{eqnarray}
&&\frac{\bar{P}+\frac32\bar{P}^{-4}}{\sqrt{\bar{P}^2-\bar{P}^{-3}}}
+\frac{3\alpha+1}{\alpha-1}\frac{\sqrt{\bar{P}^2-\bar{P}^{-3}}}{\bar{P}}
=\frac{k_6^2\alpha}{2(\alpha-1)}\ell T_4\label{OP}
\,,\\
&&\frac{\bar{P}+\frac32\bar{P}^{-4}}{\sqrt{\bar{P}^2-\bar{P}^{-3}}}
- \frac{\sqrt{\bar{P}^2-\bar{P}^{-3}}}{\bar{P}}
=\frac{\ell k_6^2\alpha \tau_4}{2K^\alpha(\bar{P}^2-\bar{P}^{-3})^{\alpha/2}}
\label{casimir}
\,.
\end{eqnarray}
At first sight these equations appear intractable for an analytic
solution, but if we first square both sides of eq.\ (\ref{OP}), we
find that it becomes quadratic in $\bar P^5$:
\be
{(64-c)\alpha^2}\bar{P}^{10}+
\left({c\alpha^2-16\alpha(3\alpha+5)}\right)\bar{P}^5+
{(3\alpha+5)^2}=0
\ee
where we defined $c=(k_6^2\ell T_4)^2$.  The solutions are given by
\be
\bar P^5_\pm = {80+\alpha(48-c) \pm\sqrt{\Delta}\over 2\alpha(64-c)},
\qquad \Delta = c(100 - 40\alpha +(c-60)\alpha^2)
\ee
From the discriminant $\Delta$ of the quadratic equation, it is
necessary that $c > c_{\rm min} \equiv 60 + 40/\alpha - 100/\alpha^2 = 
64-({10}/{\alpha}-2)^2$ to have real solutions for
$\bar P$. Furthermore, physical solutions must have $\bar P >
\bar\varrho = 1$, where from eq.\ (\ref{rhoeq}) the static position of
the 3-brane is $\bar\varrho=1$.  We find that it is only possible to
have two physical solutions for $\bar P$ in the cases where
$\alpha=2,3,4$.  For these, the second solution becomes unphysical
if $c>64$.  Thus only for the cases with $2\le\alpha\le 4$ and
$c_{\rm min} < c < 64$ do we get two possible static solutions to the
Einstein equations for the same input parameters.  For $\alpha=1$,
$\Delta = c^2$ and the solution $\bar P_-=1$ is spurious; it 
does not solve the original unsquared eq.\ (\ref{rhoeq}).
For $\alpha=1$, $c$ is restricted to the interval $[0,64]$ 
for the valid solution $\bar P_+$ to exist, since $\bar P_+<0$
for $c>64$.  For $\alpha=5$, 
$c_{\rm min}=64$ and the solution $\bar P_-$ is always negative.
For $\alpha>5$, although $c_{\rm min}<64$, $\bar P_-$ is still
negative and there is only one physical solution.
We illustrate the three qualitatively different cases by graphing 
$\bar P$ versus $c$ for $\alpha=1,3,5$ in figure \ref{pbar}. 

We can also solve eq. (\ref{casimir}) for the 3-brane tension 
in terms of $\bar P$.  To get a simpler-looking result, it is 
helpful to
once again square both sides of the equation before solving.
We find that
\be
	 {k_6^2\bar\tau_3\over 2\pi} = 
	1 - \frac52\left(\alpha \tau_4 k_6^2\over 5
l^{\alpha-1}\right)^{1/\alpha}
	(\bar P^5 - 1)^{1-\alpha\over 2\alpha} \bar P^{5+3\alpha\over
	2\alpha}
\label{taubar}
\ee
Recall that $k_6^2\bar\tau_3$ is the deficit angle of the conical
singularity at the 3-brane, which cannot exceed $2\pi$.  Eq.\
(\ref{taubar}) is consistent with this requirement, so long as
$\tau_4>0$ and $\bar P>1$.  But we should
further demand that the brane tension be positive to avoid
instabilities associated with negative tension branes.  Eq.\
(\ref{taubar}) shows that this problem can always be avoided by
choosing the free parameter $\tau_4$, quantifying the amount of
extra nontensional matter on the 4-brane, to be small enough.

\section{Modified Friedmann equation}
\label{Feq}

In this section we will use approximate and numerical methods to
solve the jump conditions (\ref{JC2:4-brane:1}, \ref{JC2:4-brane:2})
and thereby deduce the form of the Friedmann equation in the 6D
model. Before solving them, it is important to understand which
quantities should be considered as inputs and which are derived. 
Obviously we can freely specify all sources of stress energy,
including $\Lambda_6$, $T_4$, $\tau_3$, $\tau_4$ and $\alpha$; these
determine the Hubble parameter $H$.  However our de Sitter brane
solutions only exist for special values of $P$, the position of the
4-brane, which in the absence of the bulk scalar field is an
unstabilized modulus, except in the case where $\alpha>5$
\cite{Burgess:2001bn}.  Therefore eqs.\ (\ref{JC2:4-brane:1},
\ref{JC2:4-brane:2}) should be seen as determining $H$ and $P$ given
arbitrary sources of stress-energy.

We will solve the jump conditions at first ignoring  the bulk scalar
field. The Friedmann equation is obtained by finding the dependence
of $H$ on the 3-brane tension $\tau_3$ while holding other sources
fixed.  As shown above,  for given values of the 4-brane stress
energy, there exists a special value $\bar\tau_3$ of the 3-brane
tension  (and possibly two such values) which leads to the static
solution, $H=0$.  The analog of the Friedmann equation is the
functional dependence of $H$ on the difference $\delta\tau = \tau_3
-\bar\tau_3$ which gives rise to expansion of the brane.  Although
this is not a realistic situation, since we are not allowing for
radiation or matter on the 3-brane, one expects a pure tension source
to nevertheless reveal the functional dependence of the Hubble
expansion on the excess energy density of the brane.  In any region
of parameter space where the conventional Friedmann equation is 
recovered, this of course has to be the case, since the Friedmann
equation depends only upon the total energy density and not upon the
equation of state.  To explicitly study brane matter with $p\neq
-\rho$, it is necessary to regularize the 3-brane by giving it
a finite thickness \cite{Vinet,regularized}.  In the present work we
avoid these complications by assuming that purely tensional energy
density is sufficient for mapping out the functional dependence of
$H(\rho)$, but it is possible that this assumption could break down
in the regions of interest, where significant deviations from the
conventional Friedmann equation occur.  This is an interesting
question for future study, but beyond the scope of this paper.

We will use two different methods to obtain the desired relation. 
First, in section \ref{PFE} we will use a perturbative approach,
treating $H$ and $\delta\tau$ as small quantities; this gives some
analytic insight into the deviations from the standard Friedmann
equation.  Second, in section \ref{NFE}, we solve the jump conditions
numerically, allowing us to probe arbitrarily large values of the
brane tension.

\subsection{Perturbative Friedmann equation} \label{PFE} A
perturbative solution for $H$ as a function of $\tau_3$ can be found
by expanding around the static solutions found in section
\ref{staticsol}.  
In that section  we noted that for $1<\alpha<5$ there are
two static solutions for a given set of stress-energies for the source
brane.  The procedure we describe applies equally to expanding around
either of these two solutions in those cases.
The strategy is to first solve
eq.~(\ref{JC2:4-brane:1}) to obtain the position of the 4-brane in
terms $H$, $P=P({H})$. Substituting $P$ into
eq.~(\ref{JC2:4-brane:2}), we then find a relation between ${H}$ and
$\tau_3$ since the deficit parameter in eq.~(\ref{JC2:4-brane:2}) is
given by $K=K({H},\tau_3)$ from eq.~(\ref{JC2:3-brane}).  The
relation can be solved for $H$ by perturbing in the excess 3-brane
tension $\delta\tau_3 = \tau_3-\bar\tau_3$, or equivalently in the
Hubble rate $H$.  Thus we start by expressing $P$ as a Taylor series
in $H$ using eq.\  (\ref{JC2:4-brane:1}); then we will eliminate $P$
from eq.\ (\ref{JC2:4-brane:2}) to obtain $H$ as a Taylor series in
$\delta\tau_3$. The Taylor expansion for $P$ is given by
\be
	P = \bar{P} +  \frac{d\bar{P}}{d{\cal H}}{\cal H} +  \frac12 \frac{d^2\bar{P}}{d{\cal H}^2}{\cal H}^2 + ...
\label{Pexp}
\ee
where we have defined the dimensionless quantity
\be
{\cal H} \equiv H^2 \ell^2
\ee
The coefficient $\frac{d\bar{P}}{d{\cal H}}$ in this expansion can be found by differentiating
eq.\ (\ref{JC2:4-brane:1}) to get 
\be
	\frac{d\bar{P}}{d{\cal H}} = 
	\frac{4\,(\alpha+1)\bar{P}^5-9\,\alpha+1}{10\,(\alpha-5)\bar{P}^5
+5\,(3\,\alpha+5)}\bar{P}^4
\ee
To relate $H$ to the 3-brane tension, 
we use (\ref{JC2:3-brane}) and (\ref{JC2:4-brane:2}) to find that 
\begin{eqnarray}
1-k_6^2\frac{\tau_3}{2\pi}=\left(\rho+\frac{3}{2\rho^4}\right)
\left(\frac{k_6^2\alpha \tau_4}{2 G(P)\,\ell^{\alpha-1}\,f^{\alpha/2}(P)}
\right)^{1/\alpha}
\label{3bt}
\end{eqnarray}
where we recall that $\rho$ depends on $H$ via eq.\ (\ref{JC2:3-brane}).
Note that $\bar\tau_3$ is defined to be the value of $\tau_3$ corresponding to $H=0$ and $P=\bar{P}$.
To find the Friedmann equation at leading order in $\delta\tau_3$, we take $\tau_3 = \bar\tau_3
+\delta\tau_3$ and
substitute $P = \bar{P} +  \frac{d\bar{P}}{d{\cal H}}{\cal H}$ into the r.h.s.\ of (\ref{3bt}).  
This gives
\begin{eqnarray}
-\frac{k_6^2}{2\pi }\,\delta\tau_3
=-\frac{2}{5}{\cal H}(\bar{P}^3-1)\left(\,1-\frac{k_6^2}{2\pi}\bar\tau_3\,\right) 
+ O({\cal H}^2).
\label{preFRW}
\end{eqnarray}
which after solving for $H^2$ becomes
\begin{eqnarray}
{H}^2
=\frac{5}{2}(\bar{P}^3-1)^{-1}\left(\,1-\frac{k_6^2}{2\pi}\tau_3\,\right)^{-1}
\frac{k_6^2}{2\pi \ell^2}\,\delta\tau_3 + O(\delta\tau^2)
\label{FRW}
\end{eqnarray}

Ignoring the terms $O(\delta\tau^2)$, eq.\ (\ref{FRW})  has the form of the standard Friedmann equation,
$H^2= (8\pi G/3)\rho$, since $\delta \tau_3$ is the excess energy density driving the
cosmological expansion.  However we need to check that the coefficient of $\delta \tau_3$
in eq.\ (\ref{FRW}) is indeed $8\pi G/3$.  We can compute Newton's constant, or equivalently the
4D Planck mass $M_4^2 = (8\pi G)^{-1}$, by dimensionally reducing the 6D Einstein-Hilbert
action. Ignoring Kaluza-Klein excitations, the 6D and 4D Ricci scalars are related by
$R_6 = R_4/a$; thus 
\begin{eqnarray}
M_4^2 = (8\pi G)^{-1} &=& M_6^4 \int \sqrt{-g_6}\, a^{-1} dr\,d\theta =  M_6^4 \int
aK\sqrt{f}\frac{\ell}{\sqrt{f}}\, dr\,d\theta  \nonumber\\
&=&2\pi M_6^4 K \ell \int_{\varrho}^{P}\, r^2 \,dr=
\frac{2\pi}{3}M^4_6 \ell^2 
\frac{P^3-\varrho^3}{\varrho+\frac{3}{2\varrho^4}}
\left(1-k_6^2\frac{\tau_3}{2\pi}\right)
\label{graviconstant1}
\end{eqnarray}
Notice that in the above expression, the 4D Planck mass depends upon
$H$ through the integration limits $\varrho$ and $P$; thus Newton's
constant becomes time-dependent when the rate of expansion of the
universe is not constant.  We will need to check whether this effect
can be small enough to be consistent with experimental constraints
on the time variation of $G$ (see next subsection).  But for the immediate purpose of
comparing our obtained Friedmann equation with the standard one,
we define a static Planck mass which corresponds to its value
when $H=0$:
\begin{eqnarray}
\label{MP4}
\bar{M}_4^{2}= (8\pi \bar{G})^{-1} &=& 2\pi M_6^4 K\ell \int_{1}^{\bar{P}}\, r^2 \,dr = {4\pi\over 15} \ell^2 M_6^4(\bar{P}^3-1)
\left(\,1-\frac{k_6^2}{2\pi}\tau_3\,\right)
\end{eqnarray}
Since $M_6^4=
k_6^{-2}$, we see that indeed eq.\ (\ref{FRW}) is consistent with the 4D Friedmann
equation, $H^2= (8\pi \bar G/3)\delta\tau_3$.  This is in contrast to ref.\ \cite{CDGV}, which
mistakenly found deviations from the normal Friedmann equation at this order.  The mistake
made there was that corrections to the expansion rate due to changes in the 3-brane and
4-brane positions were argued to be negligible.  However, we have just shown that it is 
essential to take account of the changes in $\varrho$ and $P$ in order to obtain the correct
result.  

We emphasize that this result holds equally in the vicinity of 
either of the two static solutions that exist in the cases of $1
<\alpha < 5$. This might at first seem contradictory since the
slopes  of  $H^2$ versus $\delta\tau_3$ differ between the two
solutions.  However once one fixes the value of the 4D Planck mass by
appropriately choosing the values of the input stress-energies,
either solution equally well reproduces the standard Friedmann
equation at small values of $H$.

\begin{figure}[htp]
\centering{
\includegraphics[width=8cm]{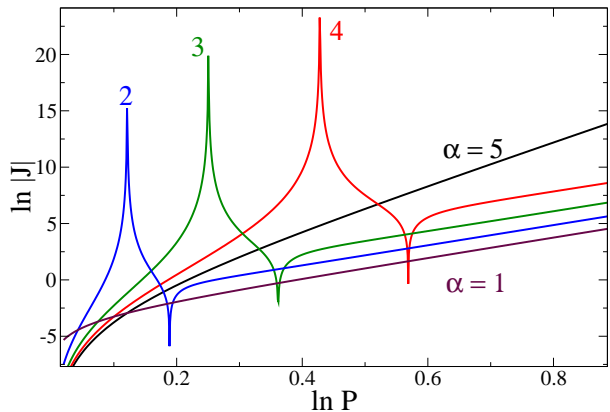}}
\caption{$\log|J|$ versus $\log \bar{P} $ for several values of $\alpha$.
$J$ is positive except for the regions between cusps.}
\label{Jfig}
\end{figure}

Having established that the Friedmann equation is recovered at leading order in $\delta\tau_3$, we now turn to the corrections at higher order. The detailed 
calculations are presented in the appendix.  
Using eq.\ (\ref{MP4}) and solving for $H^2$, we obtain
\begin{eqnarray}
H^2=\frac{8\pi \bar{G}}{3}\delta \tau_3\left(1 +
\left(\frac{8\pi \bar{G}}{3}\right)^2J\ell^4\delta\tau_3^2\right)+O(\delta\tau_3^4)
\label{FRW35}
\end{eqnarray}
where $J$ depends on $\bar P$ and $\alpha$ as shown in figure
\ref{Jfig}.
At large $\bar{P}$, $J$ increases like $\bar{P}^{6}$ for generic values of $\alpha$
and like $\bar{P}^{21}$ for $\alpha=5$ due to the behavior of the 
denominator in eq.\ (\ref{Jeq}).  
For intermediate values of  $\bar{P}$, and
$1<\alpha<5$, $J$ can become negative, as seen between the cusps 
of $\ln|J|$ in 
figure \ref{Jfig}.

\subsection{Time and spatial variation of Newton's constant}
\label{TNC}

From eq.\ (\ref{graviconstant1}) we see that the 4D Planck mass
depends on $H$ and $\tau_3$, the 3-brane tension.  We can therefore
anticipate that it will depend on time in a situation where the pure
tension $\tau_3$ is replaced by matter or radiation which gets
diluted by the expansion of the universe.  To quantify this
dependence, we will consider the variation of $M_4^2$ (or
equivalently Newton's constant $G$) with $\delta\tau_3$.  

For small $\delta\tau_3$ we can differentiate eq.\ (\ref{graviconstant1}) with respect to
$\tau_3$, keeping in mind that $P$ and $\varrho$ depend on ${\cal H}$, which in turns depends
on $\tau_3$.  At leading order, we find that
\be
{d\ln G\over d\delta\tau_3} = - {k\over\left( {\bar{P}}^{3}-1
 \right) ^{2} \left( 1-k\bar\tau_3 \right)}\left( \frac{3}{2}\,{\frac {{\bar{P}}^{6} \left( 4\, \left( \alpha
+1 \right) {\bar{P}}^{5}-9\,\alpha+1 \right) }{2\, \left( \alpha-5 \right)
{\bar{P}}^{5}+3\,\alpha+5}}+\frac32- {\bar{P}}^{3}(\bar{P}^3-1) \right)
\ee
at $\delta\tau_3=0$, where $k^2=k_6^2/(2\pi)$.  Since the deficit angle
should lie in the physical  range $[0,2\pi)$, $k^2\bar\tau_3$ must be
in $[0,1)$ which means the energy scale determining the  3-brane
tension must be approximately equal to or below the 6D Planck
mass. In the limit of large warping, $\bar{P}\gg 1$, this simplifies
to
\be
{d\ln G\over d\delta\tau_3} \cong 2 \,{4+\alpha\over 5-\alpha}\,
{k^2\over\left(  1-k\bar\tau_3 \right)}
\ee
assuming that $\alpha\neq 5$.   Let us compare this to experimental
constraints on the time-variation of $G$,  $|d\ln G/dt| < 10^{-12}/$y
\cite{Williams:2004qb}.  Since $d\rho/dt = -2\rho_0t_0^2/t^3$,  
taking
$\rho_0$ to be the present value $\sim (3\times 10^{-3}$ eV$)^4$
and $t = t_0 \sim 15$ Gyr, 
we find that the constraint implies that
\be
\label{taulimit}
{d\ln G\over d\delta\tau_3}<  {d\ln G/dt\over d\rho/dt} \cong
10^{44} {\ \rm GeV}^{-4}
\ee
Therefore $k^2\sim M_6^{-4} \lsim 10^{44} {\ \rm GeV}^{-4}$, and
we have only the extremely weak constraint that
$M_6 \gsim 10^{-11}$ GeV.

A stronger constraint can be obtained if we make the reasonable
assumption that the 6D compactification volume and hence $G$ would
be locally perturbed by a high density of matter, similarly to the
global perturbation of $G$ due to the 3-brane tension.  For example,
$G$ should not vary greatly from its known value in the vicinity of
a neutron star, in order to maintain precision tests of general
relativity from binary pulsars.  From this we conclude that $k$
should be less than the inverse density of nuclear matter, leading
to the much more interesting constraint 
\be
	M_6 \gsim 1\hbox{\ GeV}
\ee

For the special case $\alpha=5$, the above approximations are not
valid, and the time variation of $G$ is relatively large if the
radial size of the extra dimension $P$ is much greater than unity.  
We find that
\be
{d\ln G\over d\delta\tau_3} \cong  -{9\over 5}\, \bar{P}^5 \,
{k^2\over\left(  1-k^2\bar\tau_3 \right)}
\label{taulimit5}
\ee
The constraint on $M_6$ is more stringent in that case by a
factor of  $\bar{P}^5$. For the interesting value of $\bar P\sim 10^{16}$
which is required to solve the weak scale hierarchy problem, the 
constraint is 
$k^2 \lsim10^{-36}$ GeV$^{-4}$, which gives 
$M_6\geq10^{9}$ GeV, from the time variation of Newton's constant;
the corresponding limit from the spatial variation is that
$M_6$ must be close to the Planck scale.  However if the warping is weak, $\bar P\sim
1$, we retain the previous 
constraint (\ref{taulimit}).

\subsection{Numerical Results}
\label{NFE}

We can determine the relation between $H$ and $\tau_3$ numerically
using a simple algorithm. Notice that from eq.\
(\ref{JC2:4-brane:1}), one can solve for the combination $\sqrt{x}
\equiv \sqrt{P^2 - P^{-3} + H^2\varrho^2\ell^2}$ in terms of $P$ and
other known quantities, since eq.\ (\ref{JC2:4-brane:1}) is quadratic
in $\sqrt{x}$.   Then  $z\equiv (H\varrho\ell)^2 = x + 1/P^3 - P^2$. 
Furthermore, we can solve for $\varrho$ as a function of $z$ using eq.\
(\ref{rhoeq}), which can be written in the form $\varrho =  (1 +
z/\varrho^2)^{-1/5}$.  Although this has no analytic solution,
numerically it converges very quickly to the exact result, starting
from the initial guess $\varrho=1$.  At this point, we have determined
$H(P)$ since $H = \sqrt{z}/(\ell\varrho)$.  The final step is to solve
eq.\ (\ref{JC2:4-brane:2}) for $\tau_3$, which appears in the factor
$K$ through eq.\ (\ref{JC2:3-brane}).  Finally we have $H(P)$ and
$\tau_3(P)$, which allows us to plot the relation between $H$ and
$\tau_3$.

\begin{figure}[h]
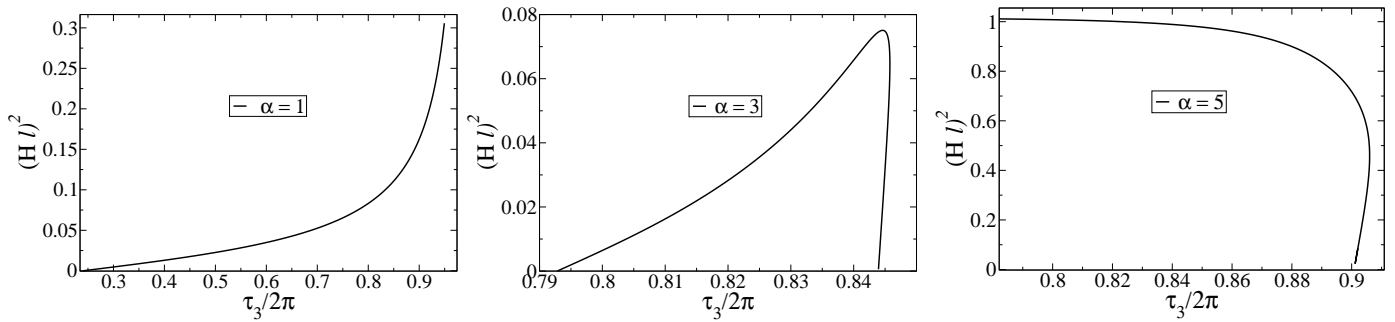

\centerline{
\includegraphics[width=6cm]{feq1.eps}
\includegraphics[width=6cm]{feq3.eps}$\ $
\includegraphics[width=6cm]{feq5.eps}}
\caption{$(H\ell)^2$ versus fractional deficit angle 
(proportional to 3-brane tension in 6D Planck units) for $\alpha=
1,3,5$, and other parameters $\ell=79.925$, $T_4=0.1$, 
$\tau_4 = 0.01$ (for $\alpha=1$), $\tau_4=1$ (for $\alpha>1$).}
\label{alpha}
\end{figure}

In figure \ref{alpha} we plot $H(\tau_3)$ for the three illustrative
cases $\alpha=1,3,5$ (recall that $\alpha$ determines the equation of
state of the extra component of matter on the 4-brane, needed for
obtaining a compact solution, with $P<\infty$).  When $\alpha=1$, 
the  solution to eq.\ (\ref{JC2:4-brane:1}) allows for the position
of the 4-brane $P\to 0$ while $H\varrho\ell\sim P^{-3/2}$.  On the other
hand eq.\  (\ref{taubar}) shows that $P\to 0$ as the deficit angle
approaches $2\pi$.   The result is that $H$ diverges for finite
$\tau_3$, and this behavior is seen in figure \ref{alpha}.  We graph
the dimensionless quantity $(H\ell)^2$ versus $\tau_3/2\pi$ in units
$M_6=1$.   In these units, $\tau_3/2\pi$ is the deficit angle of the
conical defect over $2\pi$, which should therefore lie in the range
$[0,1]$.  Notice that $H^2$ is linear in $\delta\tau_3$ in the
vicinity of $H=0$ as expected from GR, and as we showed in the
perturbative treatment.  The departure from the standard 4D behavior
becomes evident once the deficit angle starts to become large.
We used the parameter values $\ell=79.925$, $T_4=0.1$, $\tau_4 = 0.01$
in $M_6=1$ units for the $\alpha=1$ graph.  Recall from section
\ref{staticsol} that the parameter $c = (k_6^2\ell T_4)^2$ had to be
within a rather narrow range of values to get the interesting
solutions with two branches; this explains the special choice of
$\ell$.  

\begin{figure}[h]
\centerline{
\includegraphics[width=8cm]{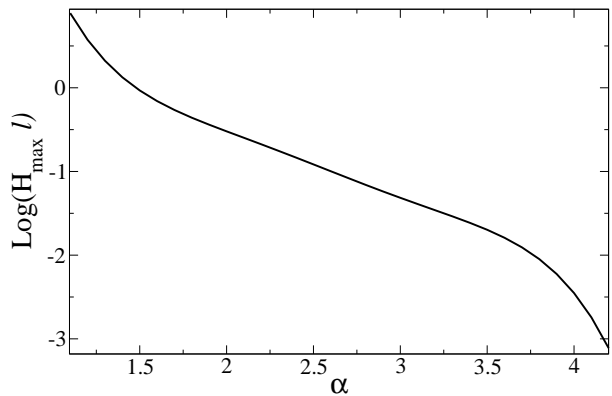}}
\caption{Maximum Hubble rate ($\log(H_{\rm max}\ell)$) as a function of $\alpha$ for
the cases similar to the middle panel of fig.\ \ref{alpha}}
\label{maxH}
\end{figure}

For the cases $1<\alpha<5$, we showed in section \ref{staticsol} that
there are two static solutions, and correspondingly,  when $H>0$
there are also two solutions for  $\sqrt{x}$.  Hence we obtain two
branches of the Friedmann equation, as shown in the second panel 
($\alpha=3$ case) of figure \ref{alpha},  but very interestingly
these two branches smoothly join with each other to make a single
curve when plotted in the $H$-$\tau_3$ plane. At small $H$, each of
the two branches is a linear function and agrees with the standard 4D
Friedmann equation, as we established in section \ref{PFE}.  The two
branches meet near a maximum allowed value of the expansion
rate, $H_{\rm max}$, whose value depends on $\alpha$.  This dependence
is shown in figure \ref{maxH}.   These examples use parameter values
$\ell=79.925$, $T_4=0.1$, $\tau_4 = 1$.

The third panel of figure \ref{alpha} illustrates the behavior when
$\alpha>5$.  Similarly to the $1<\alpha<5$ cases there is a maximum
value of $H$ (close to, but not exactly at $H\cong 1/\ell)$, but
unlike thoses cases, there is only a single branch which is connected
to a static solution.  The branch at small deficit angle asymptotes
to nonzero $H$.  The other parameter values used for this figure are
the same as for $\alpha=3$.  


\section{Friedmann equation with Goldberger-Wise scalar field}
\label{FEGW}

We have shown that the model under consideration can be consistent
with cosmological constraints on the time  and spatial variation of
Newton's constant and the standard 4D Friedmann equation at low
energies, even in the absence of a mechanism to stabilize the extra
dimensions.  However as noted in ref.\ \cite{Burgess:2001bn}, there
is an instability for $\alpha<5$ manifested by the radion having
negative mass squared, which makes it important introduce a
stabilization mechanism (and of course fifth-force constraints also
forbid a massless radion).  As in \cite{Burgess:2001bn}, we adopt the
method of Goldberger and Wise (GW), who introduced a bulk scalar
field \cite{GW,Goldberger:1999wh, Goldberger:1999un} in 5D models to
stabilize the extra dimension. Our interest will be to see how this
affects the cosmological expansion.

The basic idea behind the GW mechanism is to demand a nontrivial
background solution for a free bulk scalar field $\phi$ by including
potentials for $\phi$ on the branes which bound the bulk.  These have
the property of being minimized for nonvanishing valeus of $\phi$. 
The competition between gradient and potential energy in the bulk
causes the energy to be minimized at some intermediate value of of
the radius of the extra dimension.  In ref.\ \cite{Burgess:2001bn} it
was shown that the resulting radion mass is suppressed by an extra power
(namely $0.75$) of the warp factor relative to the 5D Randall-Sundrum
model, due to the largeness of the azimuthal extra dimension at the
Planck brane in the 6D model.  The presence of this rather large
extra dimension in fact makes the present model a kind of hybrid
between the ADD \cite{ADD} large extra dimension scenario and the
warped one.

We have given the equations of motion and boundary conditions for the
scalar field already in sections \ref{bem} and \ref{jumpcond}. 
However for the numerical methods we will use in the present section,
it is more convenient to transform to a different coordinate system,
whose line element is:
\begin{eqnarray}
ds^2=M^2(\tilde{r})[-dt^2+e^{2Ht}dx^2]+d\tilde{r}^2
+L^2(\tilde{r})\,d{\theta}^2,\label{metric2}
\end{eqnarray}
In these coordinates, the 3-brane always sits at the origin, and we
can rescale the 3-brane coordinates $x^\mu$ so that $M(0)=1$.
Defining $u=M'/M$, $v=L'/L$ \cite{CDGV}, the Einstein and scalar
field equations are
\begin{eqnarray}
\mu\mu :\quad &&v'+3u'+v^2+6u^2+3uv
=\frac{\Lambda_4}{M^2}-k_6^2\left(\frac{1}{2}\phi'^2+\cal{V}\right)\nonumber\\
\theta\theta:\quad &&4u'+10u^2
=\frac{2\Lambda_4}{M^2}-k_6^2\left(\frac{1}{2}\phi'^2+\cal{V}\right)\nonumber\\
\varrho\varrho :\quad &&4uv+6u^2
=\frac{2\Lambda_4}{M^2}+k_6^2\left(\frac{1}{2}\phi'^2-\cal{V}\right)\nonumber\\
\phi :\quad &&\phi''+(4u+v)\phi'=\frac{d\cal{V}}{d\phi}
\label{einsteinequation}
\end{eqnarray}
where we have defined $\Lambda_4 = 3 H^2$.  
In the following analysis we chose a special form for the bulk scalar
field potential
\be
\mathcal{V}=\left(\frac{b^2}{2k_6}+\frac{5b}{2l\sqrt{k_6}}\right)\phi^2-\frac{5}{32}b^2
\phi^4-\frac{10}{l^2k_6^2}
\label{bulkV}
\ee
inspired by the method of \cite{dewolfe}, in which a similar
potential was used in order to find an exact static solution for the
coupled scalar field and gravitational system.  (Note that in 6D,
$\phi$ has dimensions of mass squared and $b$ has dimensions of inverse
mass.) 
 In passing, we remark that when
$u=v$, {\it i.e.,} in the limit where $\alpha=0$ hence
$S_{\mu\nu}=S_{\theta\theta}$, an exact solution exists,
given by
\be
u=v=-\frac14 W,\quad
\phi'= {dW\over d\phi},\quad
\mathcal{V}=\frac12\left({d W\over d\phi}\right)^2-\frac58 W^2,\quad
W=\frac{b}{2}\phi^2-{\frac{4}{l\sqrt{k_6}} }
\label{spww}
\ee
Here the bulk cosmological constant term
$\Lambda_6=-{10}/{l^2k_6^2}$ has been absorbed into $\mathcal{V}$.
In this singular case where $\alpha=0$ the 4-brane  has been pushed
off to infinite radius and the extra dimensions are no longer
compactified.  Furthermore, the space does not smoothly close at
$r=0$, so the 3-brane would have to be replaced by a 4-brane with 
small radius in this solution.  We will confine our attention to $\alpha>0$ in the 
remainder.

The boundary 
conditions at the 4-brane, located at $\tilde r = P$, are given by
\begin{eqnarray}
V_P+T_4 &=& \left(6+\frac{2}{\alpha}\right)u(P)
+ \left(2-\frac{2}{\alpha}\right)v(P)
\label{bc2}\\
\frac{\alpha \tau_4}{2L(P)^\alpha}&=&v(P)-u(P)
\label{bc3}\\
\phi'(P)&=&-\frac{1}{2}\frac{dV_P}{d\phi}
 \label{bc4}
\end{eqnarray}
where $V_P$ is the scalar potential on the 4-brane, which prevents
$\phi'(P)$ from vanishing, even with no such potential on the
3-brane. The simplest nontrivial choice, which we adopt, is
\be
V_P=-\lambda \phi
\label{gwcb}
\ee 

The boundary conditions at the 3-brane at $\tilde r=0$ are more subtle 
because of the vanishing of the $r$-$r$ metric element there,
$L(0)=0$, causing $v = L'/L$ to diverge as $1/r$.  Nevertheless,
the Einstein equations (\ref{einsteinequation}) are well behaved
as $r\to 0$ because $vu$ and $r\phi'$ are finite, and $v'+v^2=0$.
For the numerical computations, we deal with this
complication by starting the integration slightly away
from the origin, 
for example at $r_0=10^{-10}$, taking
\bea
L(r_0)&=& \left(1-\frac{k_6^2\tau_3}{2\pi}\right)r_0
\label{bc1}\\
u(r_0)&=&r_0(\Lambda_4-\mathcal{V})\\
M(r_0)&=&1\\
\phi'(r_0) &=& r_0 \frac{d\mathcal{V}}{d\phi}(r_0)
\eea
while $\phi(0)$ (or $\phi(r_0)$) is to be 
determined.\footnote{Since only the ratio $v={L'}/{L}$ appears in 
the differential equation and not $L$ or $L'$ by itself, 
for numerical convenience one is free to rescale $L$ and $L'$ such that
$L(r_0)=r_0$, and restore the factor $(1-\frac{k_6^2\tau_3}{2\pi})$
afterwards.}

The differential equations can be solved numerically using the
shooting method.  For a given value of $H$, we first make a guess
for $\phi(0)$ and integrate the equations from $\tilde r=0$ until the
value $\tilde r = P$, defined to be the point where the boundary 
condition (\ref{bc2}) is satisfied.  However the condition
(\ref{bc4}) will not generally be satisfied; one has to adjust
the initial condition $\phi(0)$ to achieve this (or alternatively
tune $\lambda$ given some value of $\phi(0)$).  
Using the Newton-Raphson method, the correct value of $\phi(0)$ can 
be found after several iterations.  Once the solution is found, the
corresponding 3-brane tension can be calculated 
from eqs.\ (\ref{bc1},\,\ref{bc3}), which imply
\be
\left(1-\frac{k_6^2\tau_3}{2\pi}\right)^{\alpha}=\frac{\alpha
\tau_4}{2(v(P)-u(P))L^{\alpha}_{4}}. 
\ee 
In this way we obtain the relation between $H$ and $\tau_3$ which we
interpret as the Friedmann equation.

\begin{figure}[h]
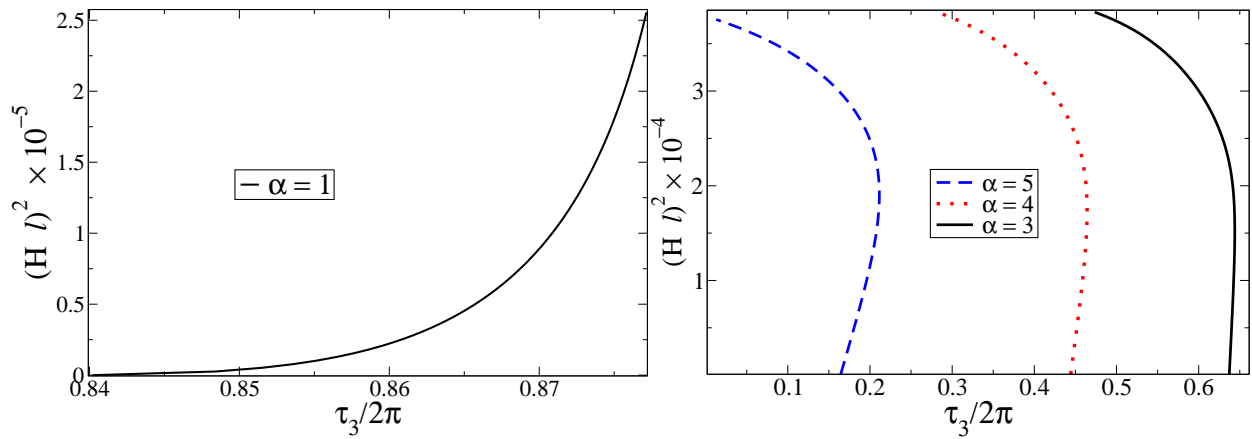

\centerline{
\includegraphics[width=8.5cm,angle=0]{FESa1.eps}
\includegraphics[width=8cm,angle=0]{FESa345.eps}}
\caption{Friedmann equation with stabilization by GW scalar field 
for $\alpha=1$ (left) and $\alpha=3,4,5$ (right)}
\label{fws}
\end{figure}

We illustrate the results in figure \ref{fws},  for the cases of
$\alpha=1,3,4,5$, and other parameters given by $b=1$,
$\lambda=-4.19532$ (this was tuned so as to give $\phi(0)=4$ in the
static case),  $T_4=0.1$,  $\tau_4=4\times 10^{-5}$ and $\ell=88$.
The shapes of the curves are qualitatively similar
to their unstabilized counterparts in figure \ref{alpha}, but the
range of possible values for the Hubble parameter is dramatically
increased in the cases $\alpha > 1$.  Departure from the linear
relation  between $H^2$ and $\tau_3$ becomes significant only at much
larger values of $H$ than for the unstabilized model, as we would
expect.

\begin{figure}[h]
\begin{minipage}{8cm}
\vspace{0.95cm}
\centerline{
\includegraphics[width=8cm]{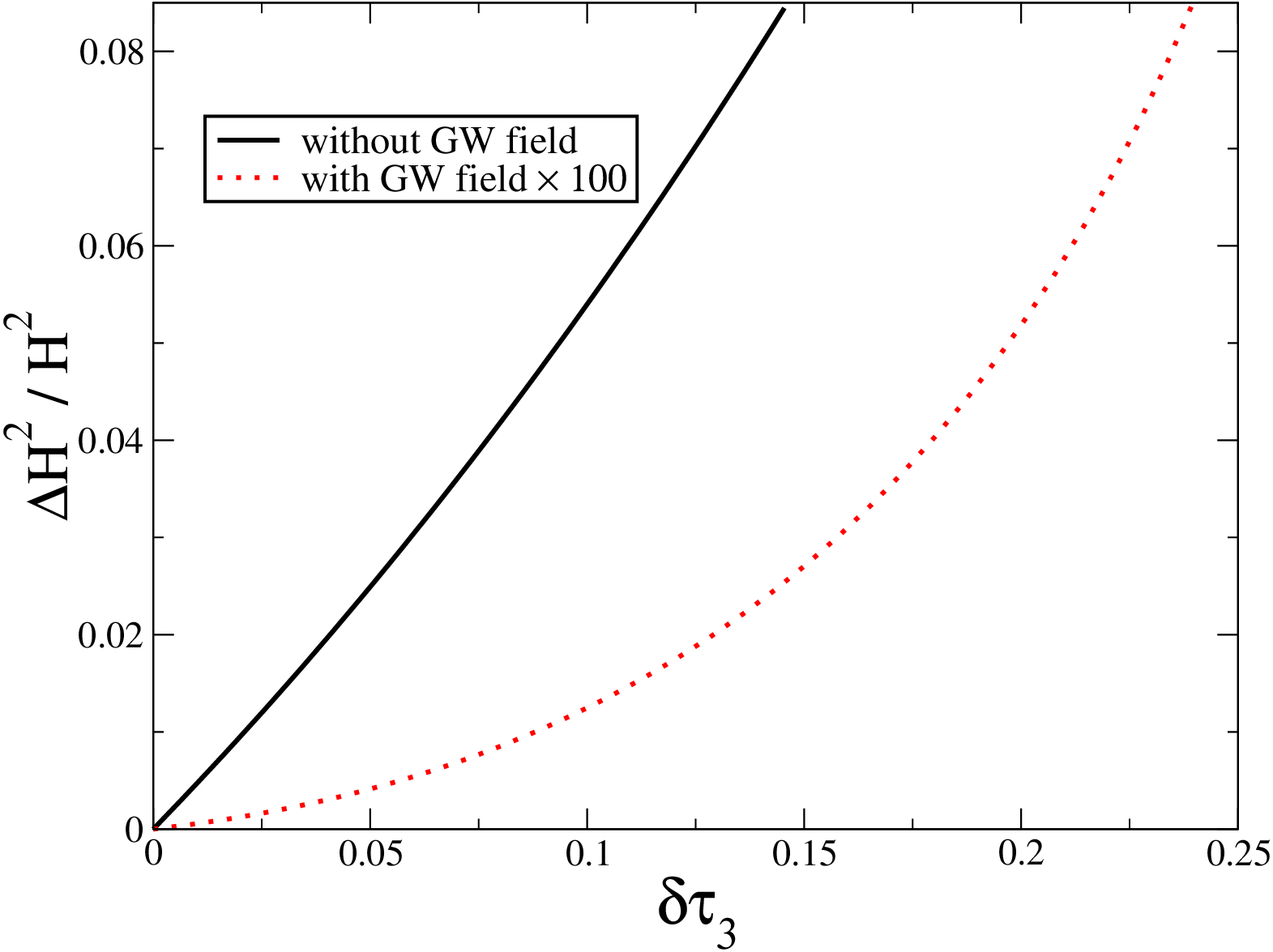}}
\caption{$\Delta H^2/{H^2}$ (defined in eq.\ (\ref{fracdev})) versus
 ${\tau_3}$ in the case with and without GW scalar field for $\alpha=5$
 }
\label{deviationwithS}
\end{minipage}
\hspace{2mm}
\begin{minipage}{8cm}
\vspace{0.95cm}
\centerline{
\includegraphics[width=8cm]{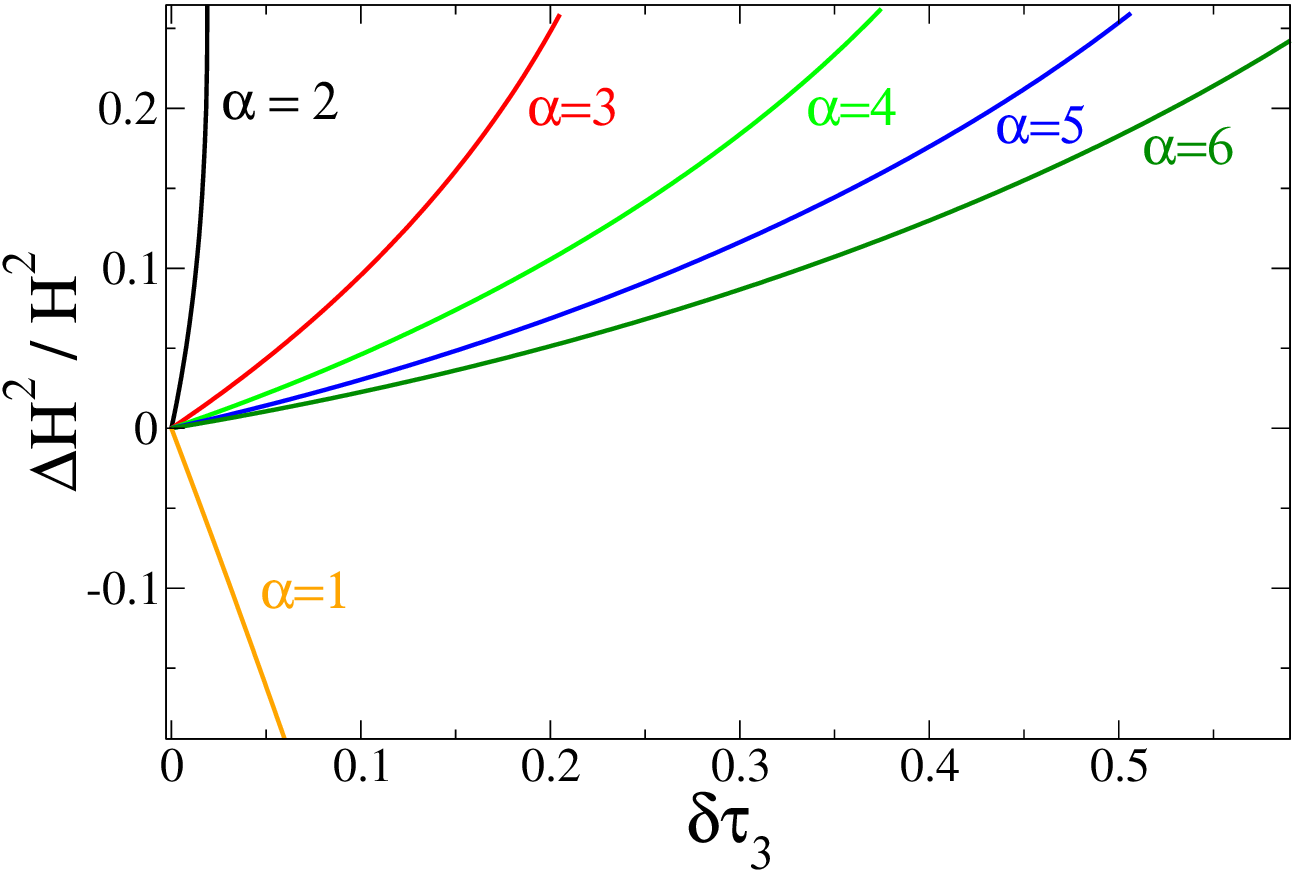}}
\caption{$\Delta H^2/{H^2}$  versus
 ${\tau_3}$ for $\alpha=1-6$ in unstabilized model.
 }
\label{fig31}
\end{minipage}
\end{figure}

To further demonstrate the effect of stabilization, we have computed
the fractional deviation of $H^2$ from its normal value, as a
function of the excess brane tension $\delta\tau_3$, for both the
stabilized and unstabilized $\alpha=5$ cases.  We define
$\Delta H^2 = H^2-{8\pi G\over  3}\delta\tau_3$, so the fractional 
deviation is 
\be
{\Delta H^2\over H^2} = 1 - {8\pi G \delta\tau_3\over 3 H^2}
\label{fracdev}
\ee
which is plotted in figure \ref{deviationwithS}.  One can see that the
deviation is more than 100 times smaller in the stabilized system
than the unstabilized one.  For reference, we also show the 
deviation in $H^2$ for other values of $\alpha$ in the unstabilized
model, in figure \ref{fig31}. We use different parameters to keep the 
4-brane position $P$ fixed for reasons that will be explained later. For 
$\alpha=6$ case the bulk is stable to some value of $H_c$ while for other 
cases the bulk is unstable for all values of $H$ which will be discussed 
in section \ref{STB}.

\begin{figure}[h]
\begin{minipage}{8cm}
\vspace{0.95cm}
\centerline{
\includegraphics[width=8cm]{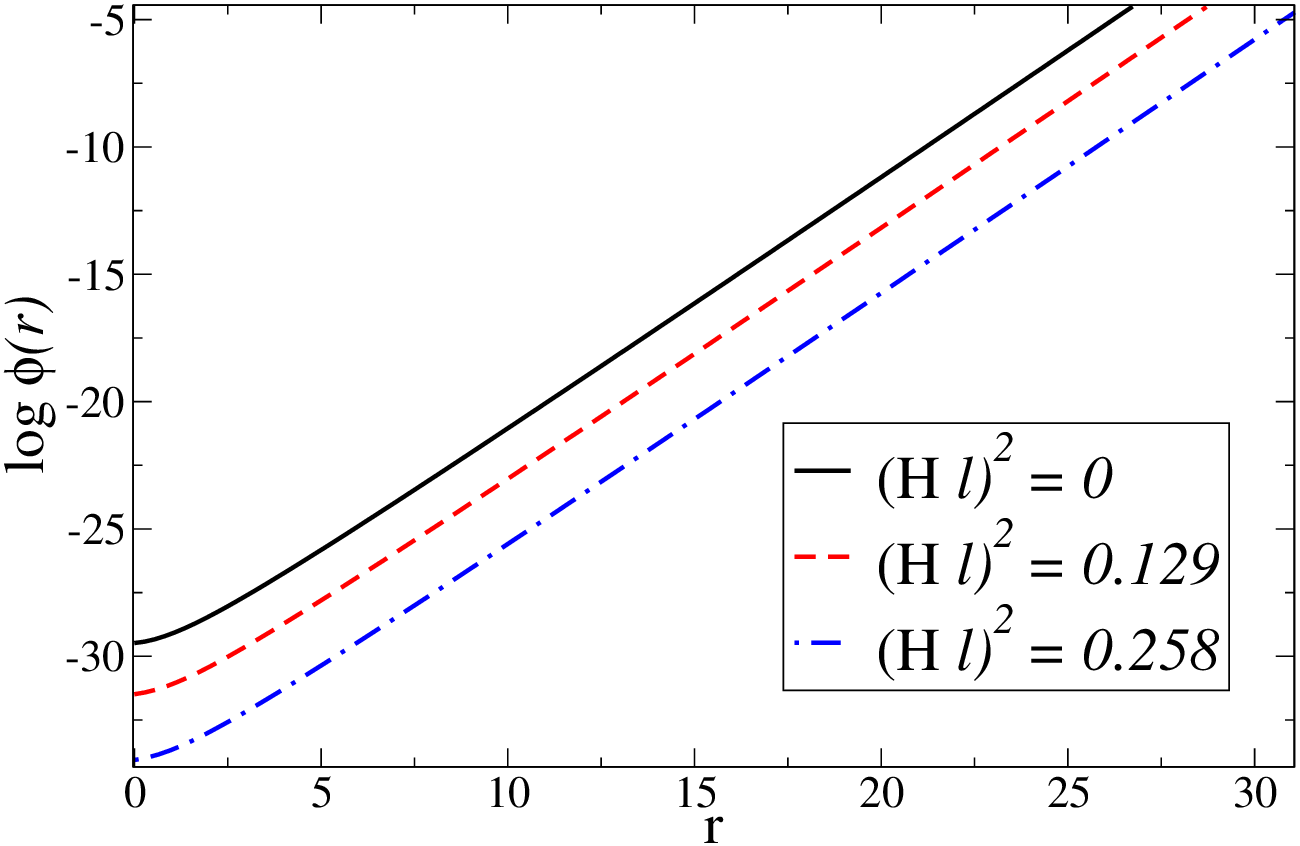}}
\caption{$\phi(r)$ for $\lambda=0.02$ and several values of $H$.
  }
\label{phirl0.01}
\end{minipage}
\hspace{2mm}
\begin{minipage}{8cm}
\vspace{0.95cm}
\centerline{
\includegraphics[width=8cm]{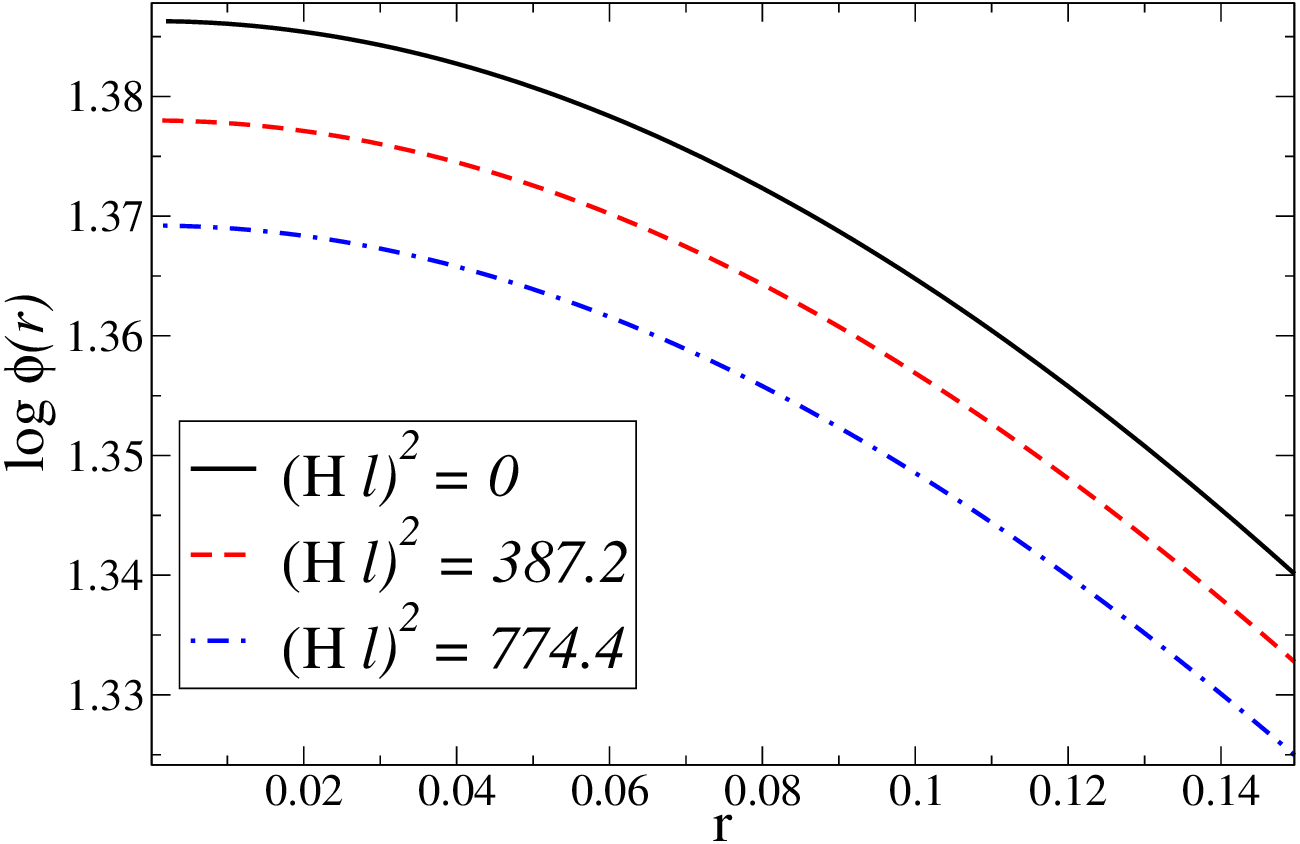}}
\caption{$\phi(r)$ for $\lambda=-4.2$ and several values of $H$. }
\label{phirl2.09}
\end{minipage}
\end{figure}

\begin{figure}[h]
\begin{minipage}{8cm}
\vspace{0.95cm}
\centerline{
\includegraphics[width=8cm]{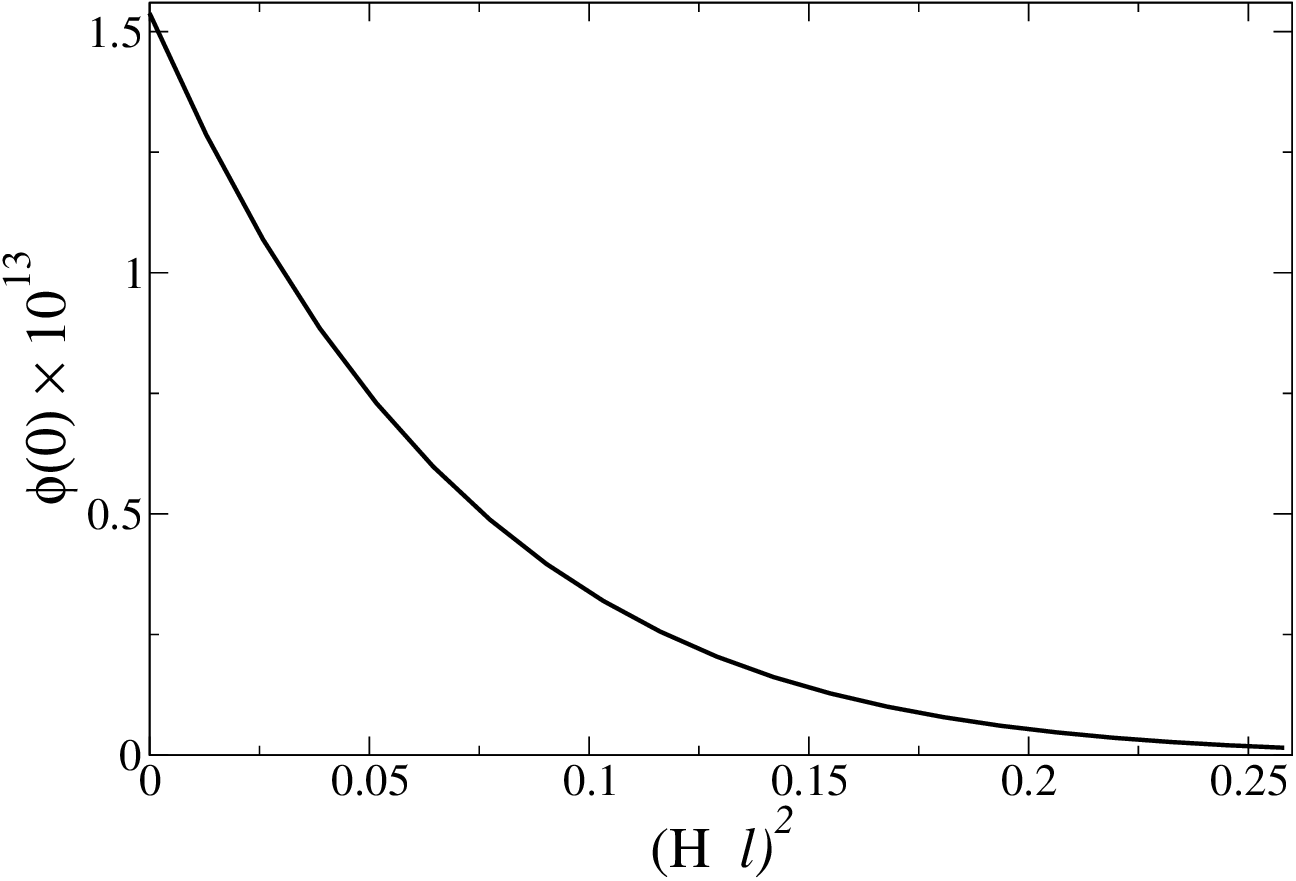}}
\caption{$\phi(0)$ for different values of Hubble rate $(H\ell)^2$, 
when $\lambda=0.02$
  }
  \label{phi0l0.01}
  \end{minipage}
  \hspace{2mm}
  \begin{minipage}{8cm}
  \vspace{0.95cm}
  \centerline{
  \includegraphics[width=8cm]{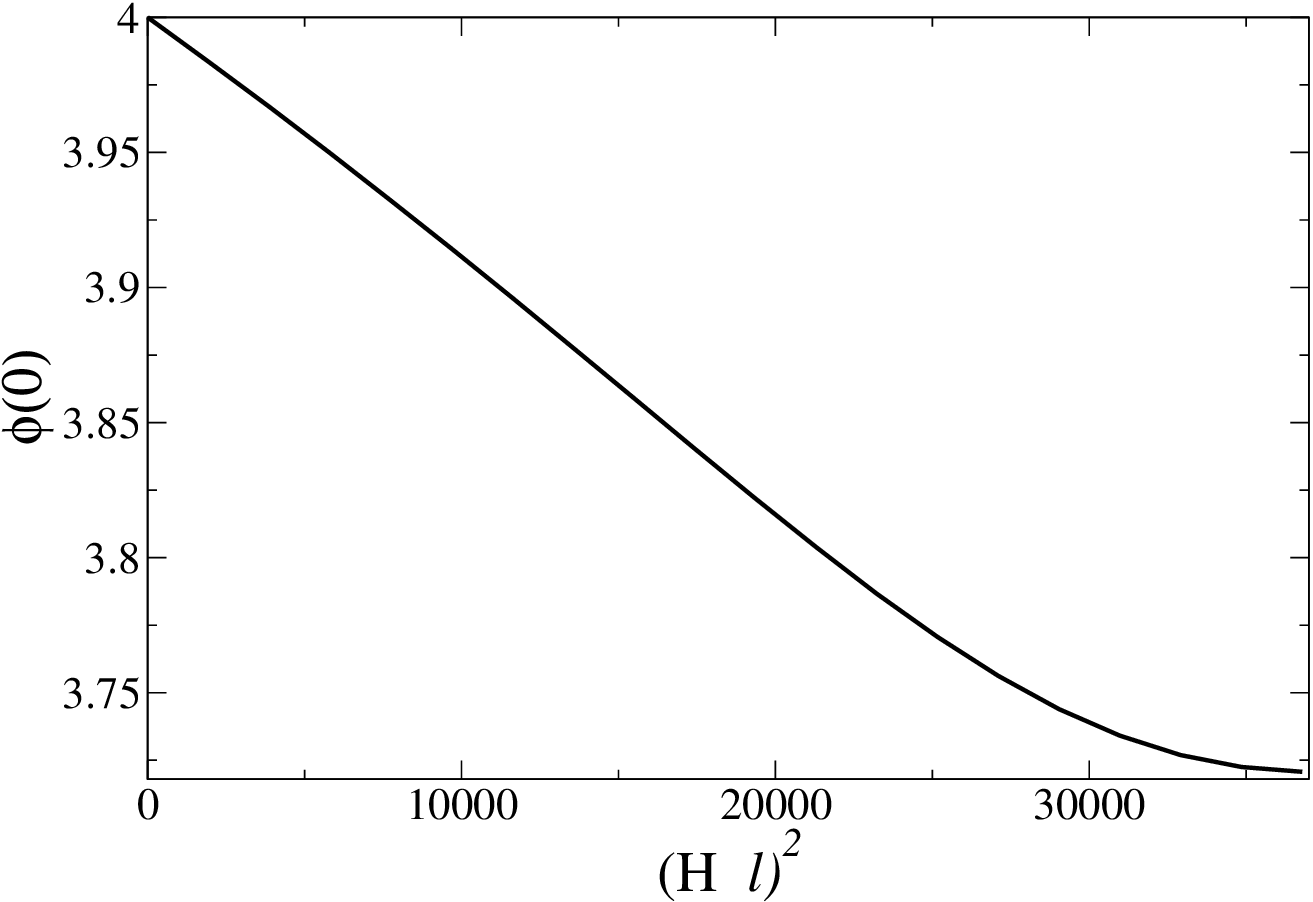}}
  \caption{$\phi(0)$ for different values of Hubble rate $(H\ell)^2$,
 when $\lambda=-4.2$    }
\label{phi0l2.09}
\end{minipage}
\end{figure}

To elucidate the behavior of the GW field we plot 
$\phi(r)$ versus $r$ for different values of the Hubble rate
and the 4-brane coupling $\lambda$ in figures 
\ref{phirl0.01} and \ref{phirl2.09}.  There are two qualitatively
different behaviors depending on whether $\lambda$ is positive or
negative; since the sign of $\phi'(r)$ at the 4-brane is governed
by $\lambda$, the two cases lead to $\phi$ either increasing or
decreasing away from $r=0$.  Moreover the magnitude of $\phi$
changes very little over the bulk if $|\lambda|$ is large, 
{\it e.g.}, $\lambda=-4.2$, 
while it varies exponentially with $r$ if $\lambda$ is small,
{\it e.g.}, $\lambda=0.02$.
The figures also show that the radial size of the extra dimension is 
roughly inversely related to the magnitude of $\lambda$.  In both
cases, the magnitude of $\phi$ is decreased when Hubble expansion is
turned on.  The dependence of $\phi(0)$ on $H$ is shown in figures
\ref{phi0l0.01}, \ref{phi0l2.09} for the two choices of $\lambda$.
Although $\phi$ decreases with $H$ in both cases, the system with
large $|\lambda|$ is much stiffer and exhibits smaller fractional
change in $\phi$. 

\section{Fluctuation analysis}
\label{STB}

To gain further insight into the issue of stability of the extra
dimensions, as it affects the cosmological expansion, we have
undertaken a stability analysis, by perturbing around the background
with arbitrary fluctuations of the metric and the GW scalar field.
Such an analysis has already been done around the static solution in 
ref.\ \cite{Burgess:2001bn}.  Here we repeat those calculations
but now we perturb around solutions with nonzero Hubble parameter.
Our goal is to quantify how the Hubble expansion affects the 
stability of the background solutions.  We will consider both
the unstabilized model, and the version which is stabilized by the
Goldberger-Wise mechanism.

\subsection{Unstabilized model}
\label{usm}
The unstable mode of our model is the radion, which is an admixture of
fluctuations of the metric functions $a,b,c$ in the ansatz
\begin{eqnarray}
ds^2=a(r,t)
\left[\,-dt^2+e^{2\tilde{H}t}\delta_{ij}dx^{i}dx^{j}\,\right]
+b(r,t)\,d\theta^2
+c(r,t)\,dr^2\,,
\label{sfluc}
\end{eqnarray}
The $(tt)+(ii)$, $(rr)$, $(\theta\theta)$ and $(tr)$ components
of the Einstein equations are
\begin{eqnarray}
&&\hspace{-5mm}
2\frac{\ddot{a}}{a}+\frac{\ddot{b}}{b}+\frac{\ddot{c}}{c}
-3\left(\frac{\dot{a}}{a}\right)^2
-\frac{1}{2}\left(\frac{\dot{b}}{b}\right)^2
-\frac{1}{2}\left(\frac{\dot{c}}{c}\right)^2
-\frac{\dot{a}}{a}\left(\,\frac{\dot{b}}{b}+\frac{\dot{c}}{c}\,\right)
-\tilde{H}\left(\,2\frac{\dot{a}}{a}+\frac{\dot{b}}{b}+\frac{\dot{c}}{c}\,\right)
=0
\cr
\cr
&&\hspace{-5mm}
\frac{3}{2}\left(\frac{a'}{a}\right)^2
+\frac{a'b'}{ab}-\frac{c}{2a}
\left[\frac{\ddot{b}}{b}+3\frac{\ddot{a}}{a}
-\frac{1}{2}\left(\frac{\dot{b}}{b}\right)^2
-\frac{3}{2}\left(\frac{\dot{a}}{a}\right)^2
+\frac{\dot{a}\dot{b}}{ab}
+9\frac{\dot{a}}{a}\tilde{H}
+3\frac{\dot{b}}{b}\tilde{H}+12\tilde{H}^2\right]
=-k_6^2c\,\Lambda_6
\cr
\cr
&&\hspace{-5mm}
2\frac{a''}{a}+\frac{1}{2}\left(\frac{a'}{a}\right)^2
-\frac{a'c'}{ac}-\frac{c}{2a}
\left[\frac{\ddot{c}}{c}+3\frac{\ddot{a}}{a}
-\frac{1}{2}\left(\frac{\dot{c}}{c}\right)^2
-\frac{3}{2}\left(\frac{\dot{a}}{a}\right)^2
+\frac{\dot{a}\dot{c}}{ac}
+9\frac{\dot{a}}{a}\tilde{H}
+3\frac{\dot{c}}{a}\tilde{H}+12\tilde{H}^2
\right]
=-k_6^2c\,\Lambda_6
+V_\theta\sqrt{c}\delta\left(\tilde{r}-P\right)
\cr
\cr
&&\hspace{-5mm}
6\frac{\dot{a}'}{a}-6\frac{\dot{a}}{a}\frac{a'}{a}
-\frac{\dot{b}}{b}\frac{a'}{a}-3\frac{\dot{c}}{c}\frac{a'}{a}
+2\frac{\dot{b}'}{b}-\frac{\dot{b}}{b}\frac{b'}{b}
-\frac{\dot{c}}{c}\frac{b'}{b}
=0
\label{pws}
\end{eqnarray}
where the 4-brane stress-energy components in eq.~(\ref{stresstensor})
are now $S_{\mu\nu}\equiv-V_0\,\tilde{g}_{\mu\nu}$ and
$S_{\theta\theta}\equiv-V_\theta\,\tilde{g}_{\theta\theta}$.
The small perturbation around the background solution is
\begin{eqnarray}
\label{a0eq}
a(r,t)&=&a_0(r)(1-a_1(r,t))\ ,\qquad a_0=r^2\\
\label{b0eq}
b(r,t)&=&b_0(r)(1-b_1(r,t))\ ,\qquad\, b_0=f(r)K^2\ ,\qquad
f(r)=\ell^{-2}\left(r^2-\frac{1}{r^{3}}+\tilde{H}^2\ell^2\right)\\
\label{c0eq}
c(r,t)&=&c_0(r)(1+c_1(r,t))\ ,\qquad\, c_0=f^{-1}(r)
\end{eqnarray}
Using the ansatz
$\ddot{a}_1+3\tilde{H}\dot{a}_1=-m_r^2a_1$ and similarly for $b_1$
and $c_1$ to expand
the Einstein equations $(tt)+(ii)$, $(rr)$ and $(tr)$
to first order we get
\begin{eqnarray}
&&2a_1+b_1-c_1=0
\label{pt:hamil}\\
&&\left(3\frac{a^\prime_0}{a_0}+\frac{b^\prime_0}{b_0}\right)a^\prime_1
-\frac{a^\prime_0}{a_0}b^\prime_1
+\frac{a^\prime_0}{a_0}\left[\,\frac{3}{2}\frac{a^\prime_0}{a_0}
+\frac{b^\prime_0}{b_0}\,\right]c_1
+\frac{c_0\,m^2_r}{2a_0}(3a_1+b_1)+6\tilde{H}^2\frac{c_0}{a_0}a_1
=0\label{pt:evo}\\
&&b_1^\prime+3\,a_1^\prime
+\frac{1}{2}\left(\frac{b_0^\prime}{b_0}
-\frac{a_0^\prime}{a_0}\right)b_1
+\frac{1}{2}\left(\frac{b_0^\prime}{b_0}+3\frac{a_0^\prime}{a_0}\right)c_1
=0
\label{pt:mc}
\end{eqnarray}
For the junction conditions at $r=P$, using $Z_2$ symmetry across
the brane for the $(tt)$ and $(\theta\theta)$ components of Einstein
equation, we find
\begin{eqnarray}
&&\left[\, \left(3\frac{a^\prime}{a}
+\frac{b^\prime}{b}\right)\,\right]\bigg|_{r=P}
=k_6^2V_0\sqrt{c}\,\big|_{r=P}\,,
\label{zo1}\\
&&\frac{4\,a^\prime}{a}\,\bigg|_{r=P}
=k_6^2V_\theta\sqrt{c}\,\big|_{r=P}\,.
\label{zo2}
\end{eqnarray}
The first order perturbation of the junction conditions is
\begin{eqnarray}
&&\left[\,\frac{3a_1^\prime+b_1^\prime}{\sqrt{c_0}}
\,\right]\bigg|_{r=P}
=\left[\,-\frac{k_6^2}{2}V_0c_1-k_6^2\delta V_0
\,\right]\bigg|_{r=P}
\label{JC:pt1}\,,\\
&&\frac{4a_1^\prime}{\sqrt{c_0}}\,\bigg|_{r=P}
=\left[\,-\frac{k_6^2}{2}V_\theta\,c_1
-k_6^2\delta V_\theta
\,\right]\,\bigg|_{r=P}
\label{JC:pt2}
\end{eqnarray}
where we used eqs.\ (\ref{zo1},\ref{zo2}) to simplify their appearance.
 Furthermore,
since $L_4\propto \sqrt{b(P)}$, we can
expand $V_0$, $V_\theta$ as
\begin{eqnarray}
\label{pot1}
&&V_0=T_4+\tau_4 L^{-\alpha}_4,\qquad\qquad\quad\,\,\,
\delta V_0=-\frac{b_1}{2b_0}\alpha\,\tau_4\,L^{-\alpha}_4 \,,\\
\label{pot2}
&&V_\theta=T_4+(1-\alpha)\tau_4 L^{-\alpha}_{4},
\,\,\qquad
\delta V_{\theta}=-\frac{b_1}{2b_0}\alpha(1-\alpha)\,\tau_4\,
L^{-\alpha}_4
\end{eqnarray}
By subtracting eq. (\ref{zo2}) from (\ref{zo1}) and using
(\ref{pot1},\ref{pot2}),  we find
\begin{eqnarray}
L^{-\alpha}_4=\frac{1}{\alpha\tau_4 k_6^2\sqrt{c_0}}
\left(-\frac{a_0^\prime}{a_0}+\frac{b_0^\prime}{b_0}\right)
\end{eqnarray}
Then eqs.~(\ref{JC:pt1}) and (\ref{JC:pt2}) lead to
\begin{eqnarray}
&&\left[\,3a_1^\prime+b_1^\prime\,\right]\big|_{r=P}
=\left[\,-\left(\frac{3a_0^\prime}{2a_0}
+\frac{b_0\prime}{2b_0}\right)c_1
-\frac{1}{2}
\left(-\frac{a^\prime_0}{a_0}
+\frac{b_0^\prime}{b_0}\right)b_1
\,\right]\bigg|_{r=P}
\label{JC:pt12}\,,\\
&&a_1^\prime\,\big|_{r=P}=
\left[\,-\frac{a_0^\prime}{2a_0}c_1
+\frac{\alpha-1}{8}
\left(-\frac{a_0^\prime}{a_0}+\frac{b_0^\prime}{b_0}\right)b_1
\,\right]\bigg|_{r=P}
\label{JC:pt22}
\end{eqnarray}
Combining eqs.~(\ref{JC:pt12}) and (\ref{JC:pt22}),
we obtain the junction condition at 4-brane
\begin{eqnarray}
\left[\,b_1'-a_1'\,\right]_{r=P}=
-\frac{1}{2}\left(\frac{b^\prime_0}{b_0}
-\frac{a^\prime_0}{a_0}\right)\left(c_1+\alpha b_1\right)\,\big|_{r=P}
\label{JC:pt-P}
\end{eqnarray}
We note that there are not really two boundary conditions at the
4-brane; imposing the momentum constraint (\ref{pt:mc})
on the junction condition (\ref{JC:pt22}) leads to the same
result as eq.~(\ref{JC:pt-P}).

At the 3-brane, we impose the condition that the deficit angle is
unchanged by perturbations around the static solution. If we consider
a circle at radius $r=\varrho+\epsilon$ around the 3-brane, with
circumference $L$ and physical radius $D$, we therefore demand that
$L/D$ be invariant in the limit that $\epsilon\rightarrow 0$:
\begin{eqnarray}
\lim_{\epsilon \to 0}\delta\left(\frac{L}{D}\right)
&=&\lim_{\epsilon \to 0}\delta
\left(\frac{\int d\theta\sqrt{b}}{\int^{\varrho+\epsilon}_\varrho dr\sqrt{c}}
\right)
=-\frac{L_0}{2D_0}\left(b_1+c_1\right)\,\bigg|_{r=\varrho}=0
\end{eqnarray}
where
\begin{eqnarray}
L_0=\int d\theta\sqrt{b_0},\quad\,D_0=\int dr\sqrt{c_0}
\end{eqnarray}
In this way, we obtain the boundary condition at 3-brane
\begin{eqnarray}
\left[\,b_1+c_1\right]\,\big|_{r=\varrho}=0
\label{JC:pt-rho}\,.
\end{eqnarray}

It proves to be convenient to work with a certain linear combination
of the perturbations, 
\begin{eqnarray}
X=3a_1+b_1 \ .
\label{cbox}
\end{eqnarray}
Using the components $(t,t)+(i,i)$ and $(t,r)$ of the Einstein
equation, the variables $a_1$, $b_1$ and $c_1$ can be expressed in terms of
$X$,
\begin{eqnarray}
a_1=\frac{b_0}{2b^\prime_0}
\left[X^\prime+\left(\frac{a^\prime_0}{a_0}
+\frac{b^\prime_0}{b_0}\right)X\right]
\,,\quad
b_1=-\frac{3b_0}{2b^\prime_0}
\left[X^\prime+\left(\frac{a^\prime_0}{a_0}
+\frac{b^\prime_0}{3b_0}\right)X\right]
\,,\quad
c_1=-\frac{b_0}{2b^\prime_0}
\left[X^\prime+\left(\frac{a^\prime_0}{a_0}
-\frac{b^\prime_0}{b_0}\right)X\right]
\end{eqnarray}
where $\prime$ denotes the derivative with respect to $r$.

Now we are ready to analyze the radion mass spectrum.
Combining eqs.~(\ref{pt:hamil})-(\ref{pt:mc}), we get
\begin{eqnarray}
X^{\prime\prime}+\left[
\frac{4}{r}+2\frac{f^\prime}{f}-\frac{f^{\prime\prime}}{f^\prime}-\frac{6f}{r^2f^\prime}+\frac{6\tilde{H}^2\ell^2}{r^2f^\prime}\right]X^\prime
+\left[\frac{m^2_r\ell^2}{r^2f}-\frac{2f^{\prime\prime}}{rf^\prime}
+\frac{4f^\prime}{rf}-\frac{12f}{r^3f^\prime}+\frac{6\tilde{H}^2\ell^2}{r^2f}
+\frac{12\tilde{H}^2\ell^2}{r^3f^\prime}\right]X=0
\label{master}
\end{eqnarray}
The boundary conditions (\ref{JC:pt-P}) and (\ref{JC:pt-rho}) become
\begin{eqnarray}
&&\left[X^\prime+\frac{2}{r}X\right]\,\bigg|_{r=\varrho}=0\label{bc1-1}\\
&&\left[\,X^{\prime\prime}+
\bigg\{\frac{3\alpha+13}{8}\frac{f^\prime}{f}
-\frac{f^{\prime\prime}}{f^\prime}-\frac{3\alpha-7}{4r}
\bigg\}X^\prime
+\bigg\{\frac{\alpha+5}{2r}\frac{f^\prime}{f}+\frac{\alpha-1}{8}
\left(\frac{f^\prime}{f}\right)^2
-\frac{3\alpha+5}{2r^2}-\frac{2f^{\prime\prime}}{rf^\prime}
\bigg\}X\,\right]
\,\Bigg|_{r=P}=0
\label{bc2-1}
\end{eqnarray}
Eliminating $X^{\prime\prime}$ from eq.~(\ref{bc2-1}) using eq.~(\ref{master}),
we find
\begin{eqnarray}
&&\left[\bigg\{\frac{3(\alpha-1)}{8}\frac{f^\prime}{f}-\frac{3(\alpha+3)}{4r}
+\frac{6f}{r^2f^\prime}-\frac{6\tilde{H}^2\ell^2}{r^2f^\prime}\bigg\}X^\prime
\right.\nonumber\\
&&\qquad\left.
+\bigg\{-\frac{m^2_r\ell^2}{r^2f}+\frac{\alpha-3}{2r}\frac{f^\prime}{f}
+\frac{\alpha-1}{8}\left(\frac{f^\prime}{f}\right)^2
-\frac{3\alpha+5}{2r^2}+\frac{12f}{r^3f^\prime}
-\frac{6\tilde{H}^2\ell^2}{r^2f}-\frac{12\tilde{H}^2\ell^2}{r^3f^\prime}\bigg\}X
\right]\,\Bigg|_{r=P}=0
\label{bc2-12}
\end{eqnarray}
Thus if we solve eq.~(\ref{master}) under the boundary condition
(\ref{bc1-1}) and (\ref{bc2-12}), we can obtain the radion mass spectrum.
In the asymptotic region $r\rightarrow\infty$, we have
\begin{eqnarray}
\left[\bigg\{-\frac{3(\alpha-1)}{4r^3}\tilde{H}^2\ell^2
+\frac{15(\alpha-5)}{8r^6}\bigg\}X^\prime
+\bigg\{-\frac{m^2_r\ell^2}{r^4}-\frac{2(\alpha+1)}{r^4}\tilde{H}^2\ell^2
+\frac{10(\alpha-5)}{2r^7}\bigg\}X\right]\,\Bigg|_{r=P}=0
\end{eqnarray}
from which we obtain the simple relation between the radion mass
and the solution of bulk equation of motion.
\begin{eqnarray}
m^2_r=\bigg(\frac{5(\alpha-5)}{\ell^2r^3}-2(\alpha+1)\tilde{H}^2
+\frac{15(\alpha-5)}{8\ell^2r^2}\frac{X'}{X}
-\frac{3}{4}(\alpha-1)\tilde{H}^2\left(\frac{rX^\prime}{X}\right)
\,\bigg)\bigg|_{r=P}
\label{zl}
\end{eqnarray}
which is consistent with the result in ref.\ \cite{Burgess:2001bn} when 
$\tilde{H}\equiv H/(1+H^2\ell^2)^{(1/5)}=0$. Note that the radion mass is 
expressed in terms of $P$ so to numerically compare results with different 
values of $\alpha$ $P$ should be fixed as we did in section \ref{FEGW}. 

In ref.\ \cite{Burgess:2001bn} it was noticed that one can obtain
an analytic approximation for the radion mass, using only the 
asymptotic behavior of the solutions for $X$, which is valid in 
the case of strong warping, $P\gg 1$.  At large $r$, the differential
equation for $X$ simplifies to the form
\be
	X'' + {4\over r} X'  +{\cal O}(r^{-4}) X = 0
\ee
which has solutions of the form $X\cong c_1 + c_2/r^3$.  Therefore
generically $X'/X\sim 1/r^4$, and we can ignore the terms proportional
to $X'/X$ in eq.\ (\ref{zl}).  Applying this observation in the
present case where we take into account the Hubble expansion 
leads to the result
\begin{eqnarray}
m^2_r=\left(\frac{5(\alpha-5)}{\ell^2P^3}-2(\alpha+1)
\tilde{H}^2\right)\left(1 + {\cal O}(P^{-3})\right)
\label{radionnoGW}
\end{eqnarray}
We can notice a correlation between the radion mass
squared and the deviations from the normal Friedmann equation,
parametrized by $\Delta H^2/H^2$ in eq.\ (\ref{fracdev}).
We find that $|\Delta H^2/H^2|$ is larger the more negative $m^2_r$
is, as one would intuitively expect.  In figure \ref{fig31} we
showed $\Delta H^2/H^2$ versus $\delta\tau_3$ for $\alpha=1-6$.
The smallest deviation, in magnitude, occurs for the largest value of
$\alpha$.

The radion mass squared is always negative when $\alpha<5$, but for
$\alpha>5$, it can be  positive if the Hubble rate is below some
critical value, defined by $\tilde H^2 =
\frac52{\alpha-5\over\alpha+1}\ell^{-2}P^{-3}$.  If the warping is
large, $P\gg 1$, and the relation 
$\tilde{H}\equiv H/(1+H^2\ell^2)^{(1/5)}=0$ implies that $\tilde
H\cong H$; in that case the critical value of $H$ is
\be
	H_c \approx \left({5(\alpha-5)\over2(\alpha+1)}\right)^{1/2}
	\ell^{-1}P^{-3/2}
\label{Hc}
\ee
For the  $\alpha=5$ case, where the anisotropy  of
the 4-brane stress tensor is provided by the Casimir effect, the
leading contribution at $H=0$ vanishes, and one must look to the 
subleading terms of order $P^{-6}$.  Of course, the approximation
(\ref{Hc}) is only valid for large warping, $P\gg1$.  In the
discussion that follows, we will use the exact value of $H_c$ which
comes from solving the radion mass eigenvalue problem numerically.

\begin{figure}[h]
\begin{minipage}{8cm}
\vspace{0.95cm}
\centerline{
\includegraphics[width=8cm]{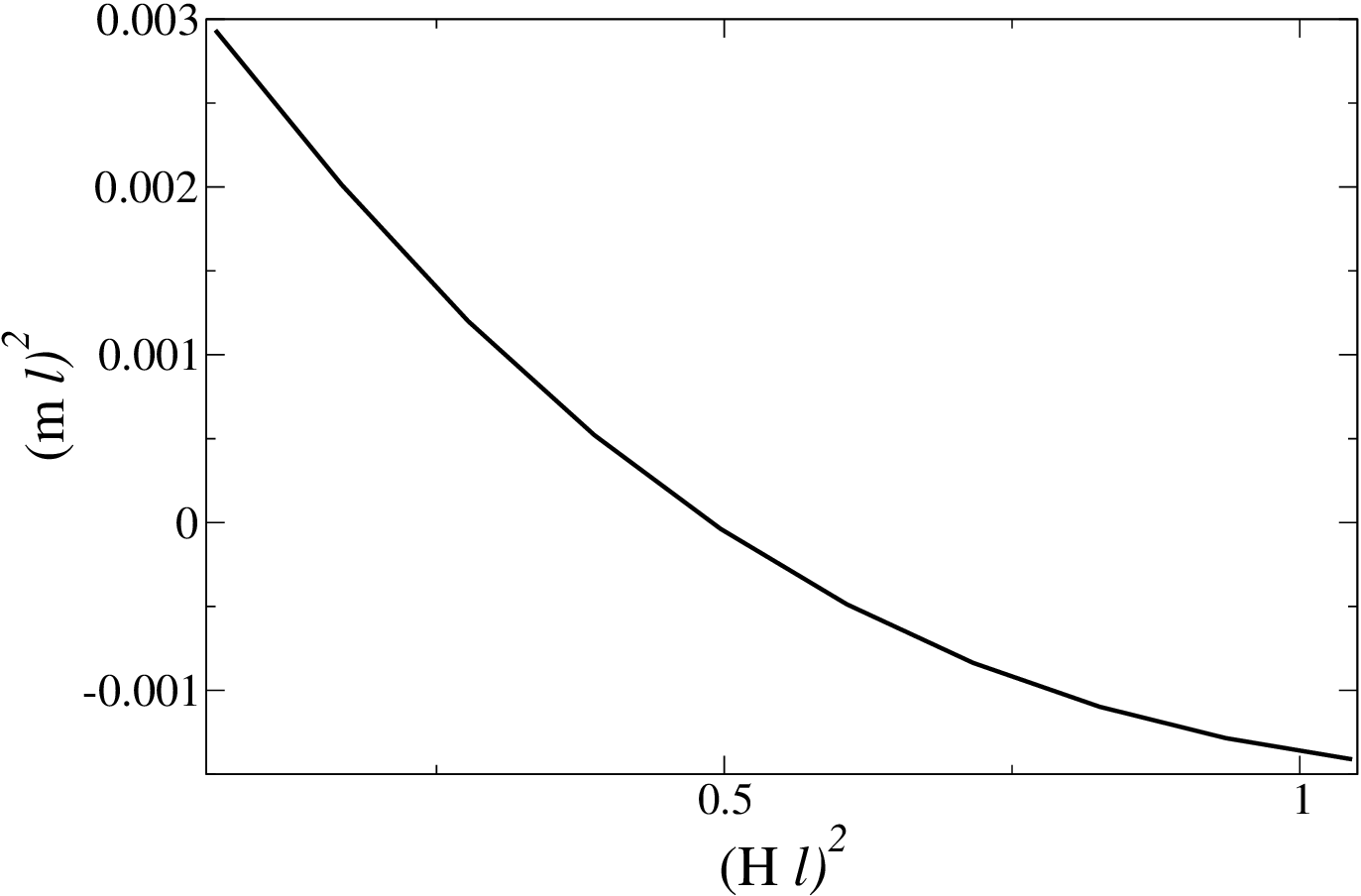}}
\caption{Radion mass squared as a function of $(H\ell)^2$ for 
$\alpha=6$.}
\label{rma6}
\end{minipage}
\hspace{2mm}
\begin{minipage}{8cm}
\vspace{0.95cm}
\centerline{
\includegraphics[width=8cm]{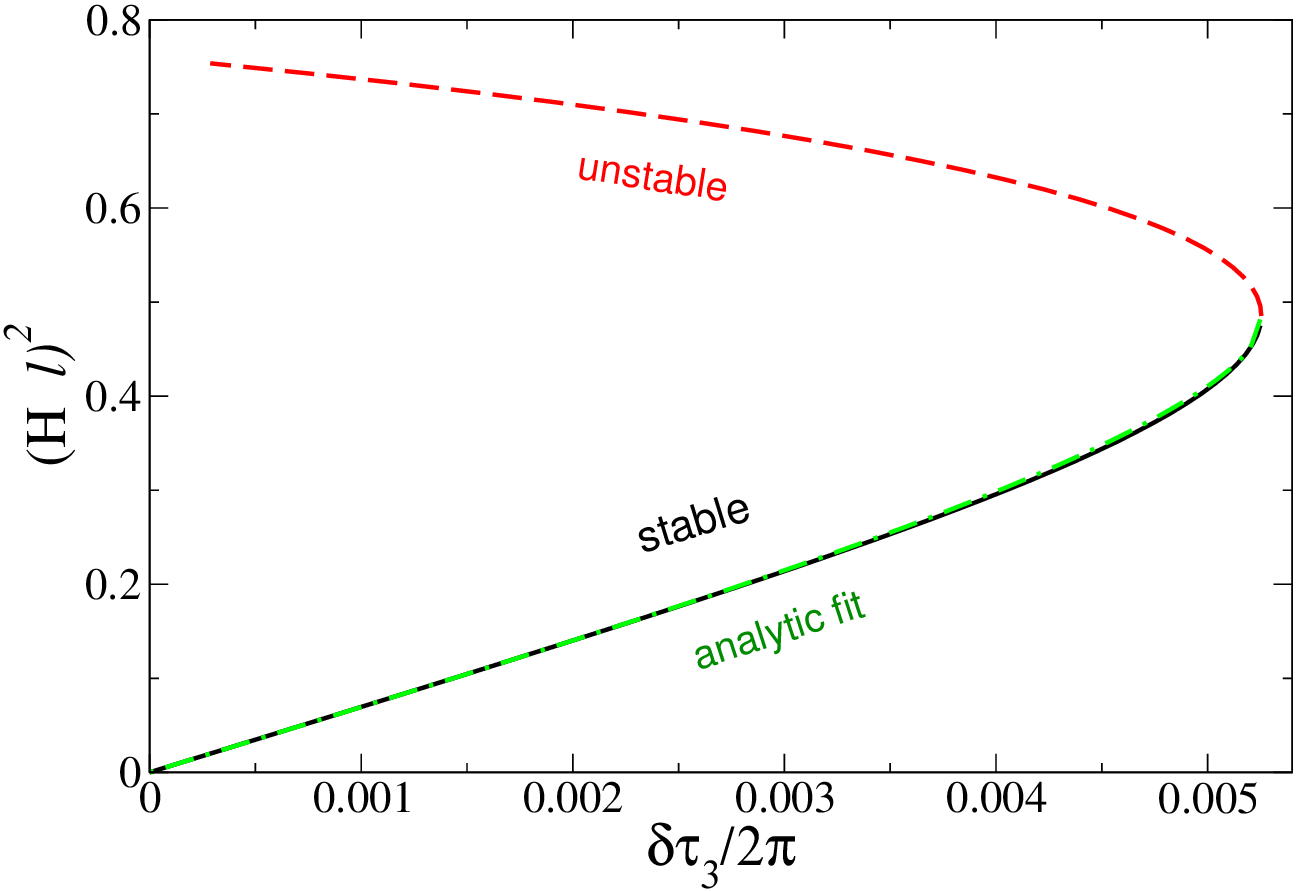}}
\caption{Friedmann equation for $\alpha=6$ showing which parts
correspond to stable and unstable radion.}
\label{fig13}
\end{minipage}
\end{figure}

In section \ref{IFL} we will make use of the interesting case of
$\alpha>5$, for which the radion is stable even without the introduction of
an external stabilization mechanism.\footnote{Unfortunately we are not
aware of any physical  models which give such an equation of state for the
extra matter on the 4-brane.}   For $\alpha>5$ we can pose the question of
how much of the exotic Friedmann relation at high values of $H$ is
consistent with stability of the radion.  To accurately address this issue,
we have numerically solved the eigenvalue problem  eq.\ (\ref{master}) for
the radion mass rather than relying upon the approximation
(\ref{radionnoGW}) or (\ref{Hc}).  The result is shown in figure \ref{rma6}, which plots
$m_r^2$ as a function of $H^2$, for $\alpha = 6$, $T_4=0.1$, $\ell=88$ and
$\tau_4=1$. The radion becomes destabilized when $(H\ell)^2\gsim 0.5$.   In
figure \ref{fig13} we show  the Friedmann relation in the vicinity of the
transition region between stable and unstable cases.  

\begin{figure}[h]
\begin{minipage}{8cm}
\vspace{0.95cm}
\centerline{
\includegraphics[width=8cm]{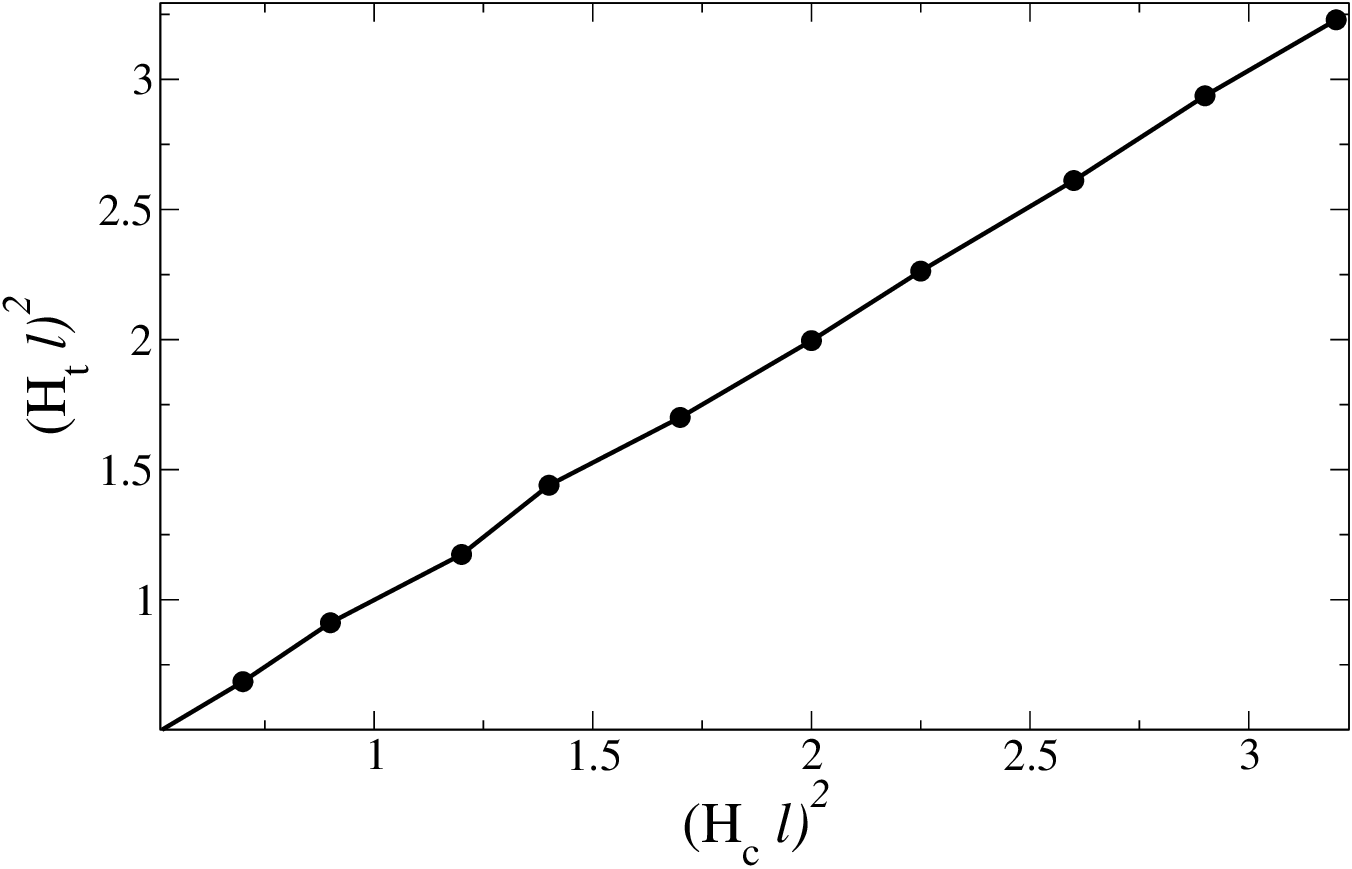}}
\caption{Hubble rate $(H_t\ell)^2$ at the turning point (where the two
branches of $H$ come together) versus $(H_c\ell)^2$, the critical
$H$ above which the radion is destabilized.}
\label{Hct}
\end{minipage}
\hspace{2mm}
\begin{minipage}{8cm}
\vspace{0.95cm}
\centerline{
\includegraphics[width=8cm]{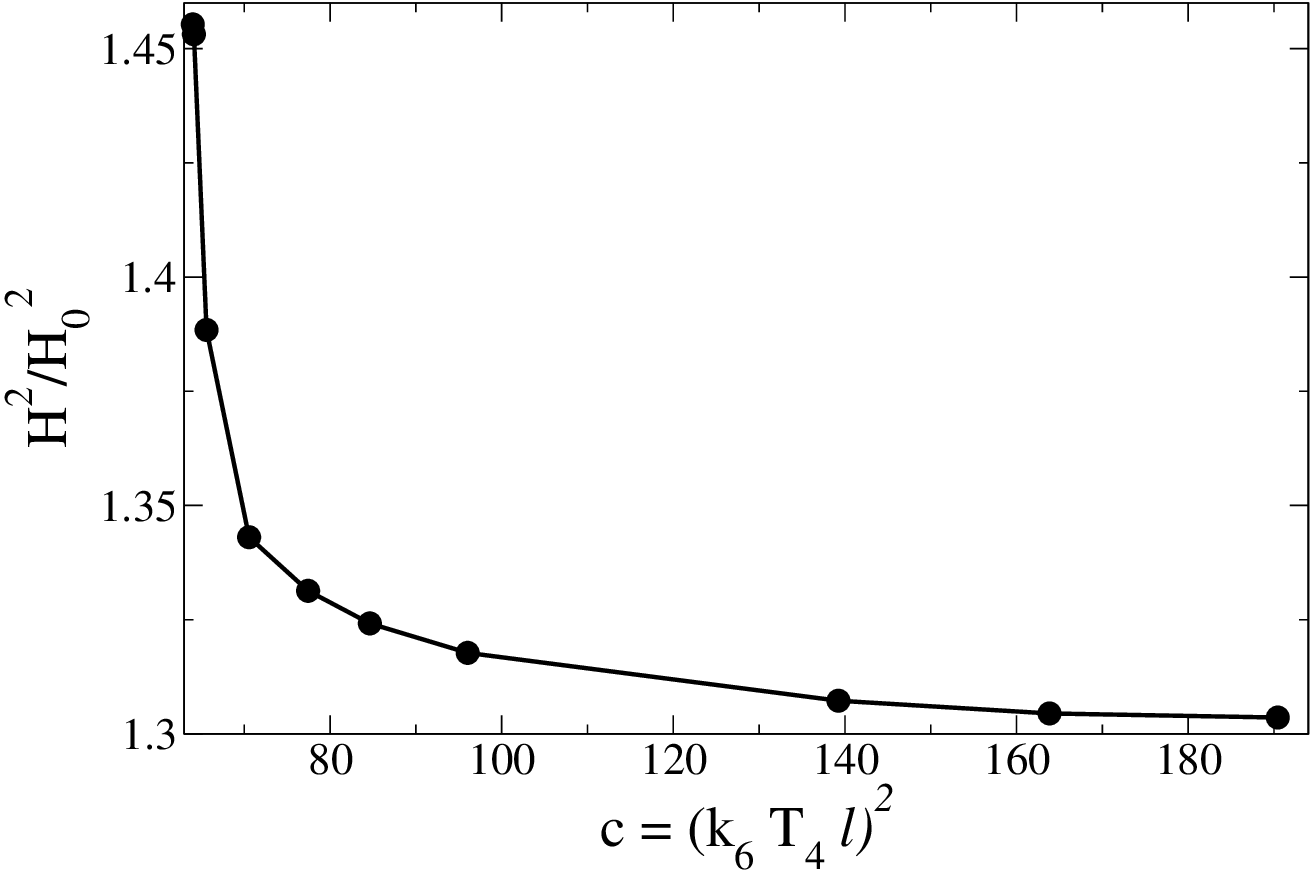}}
\caption{$H^2/H_0^2$ at the turning point for different values of $c$,
where $H_0^2= {\delta \tau_3}/{3\bar{M}_p^2}$ is the standard Friedmann
relation.}
\label{rHh}
\end{minipage}
\end{figure}

It is interesting that the transition between stable and unstable radion
occurs quite close to the joining of the two branches which make $H$
double-valued. In figure \ref{Hct} we plot the  value of  $(H_t\ell)^2$ at
the turning point as a function of  $(H_c\ell)^2$, with the curve
parametrized by varying  $c=(k_6T_4\ell)^2$.  (We still use $\tau_4=1$
here, but in fact the value of  $\tau_4$ does not affect the result because
it only appears on the right  hand of eq. (\ref{JC2:4-brane:2}) and can be
rescaled away.)  Figure \ref{Hct} shows that to within numerical accuracy,
$H_c$ and $H_t$ coincide: thus the exotic branch of the Friedmann equation,
which leads  a double-valued Hubble rate, is unstable to decompactification
of the radius of the extra dimensions.  

It is also interesting to know how strongly $H$ deviates from the
prediction of general relativity before the critical value $H_c$ is
reached.  Let us define the standard Hubble rate as $H_0^2 =  
{\delta \tau_3}/{3\bar{M}_p^2}$.  We calculated $H_c^2/H_0^2$ as a function of
$c$ and show the result in figure \ref{rHh}.  There it is seen that the
maximum deviation is no greater than approximately $1.5$ (although with
extreme fine-tuning of $c$ larger deviations might be possible).  Such
a moderate deviation might seem unimportant if it occurred early enough
in the history of the universe.  However we will show in section \ref{IFL}
that there can be interesting effects during inflation, due to the
derivative of $H$ with respect to $\tau_3$ becoming singular at the
critical point.

\subsection{Fluctuations with Goldberger-Wise stabilization}
\label{gwstab}
Carrying out the analogous steps as before, but now including
the GW bulk scalar field, we derive the coupled equations
of motion for the radion field $X$ and the fluctuations of the 
GW field, $\phi_1$.  Using the shorthand $A'_0 = a'_0/a_0$,
and similarly for $b$ and $c$, we find
\begin{eqnarray}
&&X^{''}+\left(-2A^{'}_0-B^{'}_0+\frac{3}{2}\frac{A^{'2}_0}{B^{'}_0}
-\frac{B^{''}_0}{B^{'}_0}-\frac{6c_0}{a_0}\frac{\tilde{H}^2}{B^{'}_0}
-\frac{1}{2B^{'}_0}k_6^2\phi^{'2}_0\right)X^{'}\nonumber\\
&&\quad+\left(-A_0^{''}+\frac{A^{'}_0}{B^{'}_0}B^{''}_0
-\frac{1}{2B^{'}_0}(A^{'}_0-B^{'}_0)(3A^{'2}_0+2A^{'}_0B^{'}_0)
+\frac{c_0}{a_0}m^2+\frac{6c_0}{a_0}\frac{\tilde{H}^2}{B^{'}_0}
(A^{'}_0+B^{'}_0)+\frac{1}{2B^{'}_0}(A^{'}_0-B^{'}_0)k_6^2\phi^{'2}_0\right)X
\nonumber\\
&&\quad+\left(-\phi^{''}_0+\frac{B^{''}_0}{B^{'}_0}-\frac{1}{2B^{'}_0}
(3A^{'2}_0+2A^{'}_0B^{'}_0)\phi^{'}_0
+\frac{6c_0}{a_0}\frac{\tilde{H}^2}{B^{'}_0}\phi^{'}_0
-c_0\frac{dV}{d\phi}+\frac{k_6^2}{2B^{'}_0}\phi^{'3}_0\right)2k_6^2\phi_1
=0
\label{eqX}\\
&&\phi_1^{''}-\left(2A^{'}_0+\frac{1}{2}B^{'}_0+\frac{1}{2}C^{'}_0\right)\phi^{'}_1
+\left(-c_0\frac{d^2V}{d\phi^2_0}
+\frac{c_0}{B_0}\frac{dV}{d\phi_0}k_6^2\phi^{'}_0
+\frac{c_0}{a_0}m^2\right)\phi_1\nonumber\\
&&\quad-\left(\phi^{'}_0+\frac{c_0}{2B^{'}_0}\frac{dV}{d\phi}\right)X^{'}
+\frac{c_0}{2B^{'}_0}(A^{'}_0-B^{'}_0)\frac{dV}{d\phi_0}X=0
\end{eqnarray}
The boundary conditions for $X$ are 
\begin{eqnarray}
&&X^{'}-A^{'}_0X=0|_\varrho \\
&&\left(\frac{3(\alpha+3)}{8}A^{'}_0-\frac{3(\alpha-1)}{8}B^{'}_0
-\frac{3}{2}\frac{A^{'2}_0}{B^{'}_0}
+\frac{6c_0}{a_0}\frac{\tilde{H}^2}{B^{'}_0}
+\frac{1}{2B^{'}_0}k_6^2\phi^{'2}_0\right)X^{'}\nonumber\\
&&+\left[\frac{1}{8}(B^{'}_0-A^{'}_0)((-7+3\alpha)A^{'}_0+(\alpha-1)B^{'}_0)
+\frac{3A^{'2}_0}{2B^{'}_0}(A^{'}_0-B^{'}_0)-\frac{c_0}{a_0}m^2
-\frac{6c_0}{a_0}\frac{\tilde{H}^2}{B^{'}_0}(A^{'}_0+B^{'}_0)
-\frac{1}{2B^{'}_0}(A^{'}_0-B^{'}_0)k_6^2\phi^{'2}_0\right]X\nonumber\\
&&-2k_6^2\phi^{'}_0\phi^{'}_1+\left(\frac{1+3\alpha}{8}B^{'}_0\phi^{'}_0
+\frac{7-3\alpha}{8}A^{'}_0\phi^{'}_0+\frac{3A^{'2}_0}{2B^{'}_0}\phi^{'}_0
-\frac{6c_0}{a_0}\frac{\tilde{H}^2}{B^{'}_0}\phi^{'}_0+c_0\frac{dV}{d\phi}
-\frac{k_6^2}{2B^{'}_0}\phi^{'3}_0\right)2k_6^2\phi_1=0
\label{gwbc}
\end{eqnarray}

\begin{figure}[h]
\begin{minipage}{8cm}
\vspace{0.95cm}
\centerline{
\includegraphics[width=8cm]{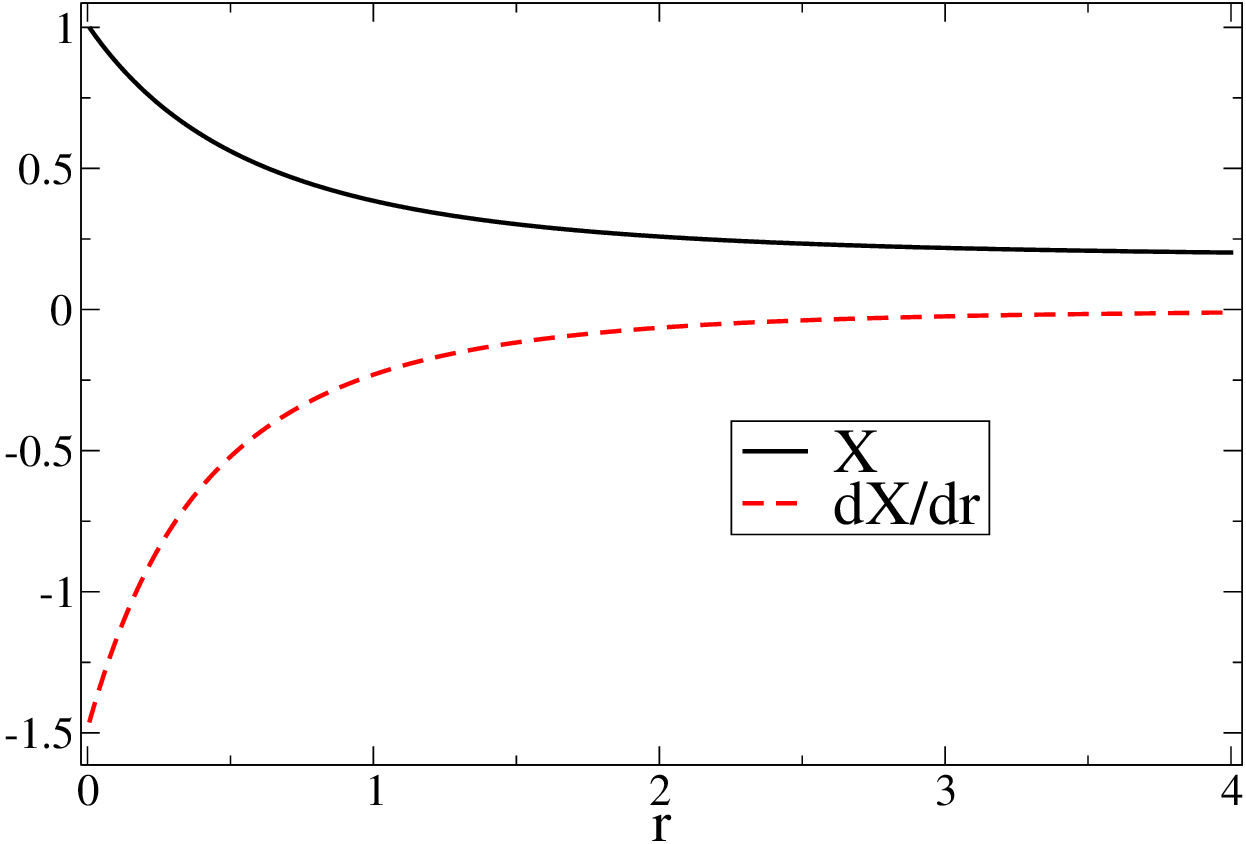}}
\caption{Radion wave function $X$ and its derivative,
 with stabilization by GW mechanism.}
\label{wavex}
\end{minipage}
\hspace{2mm}
\begin{minipage}{8cm}
\vspace{0.95cm}
\centerline{
\includegraphics[width=8cm]{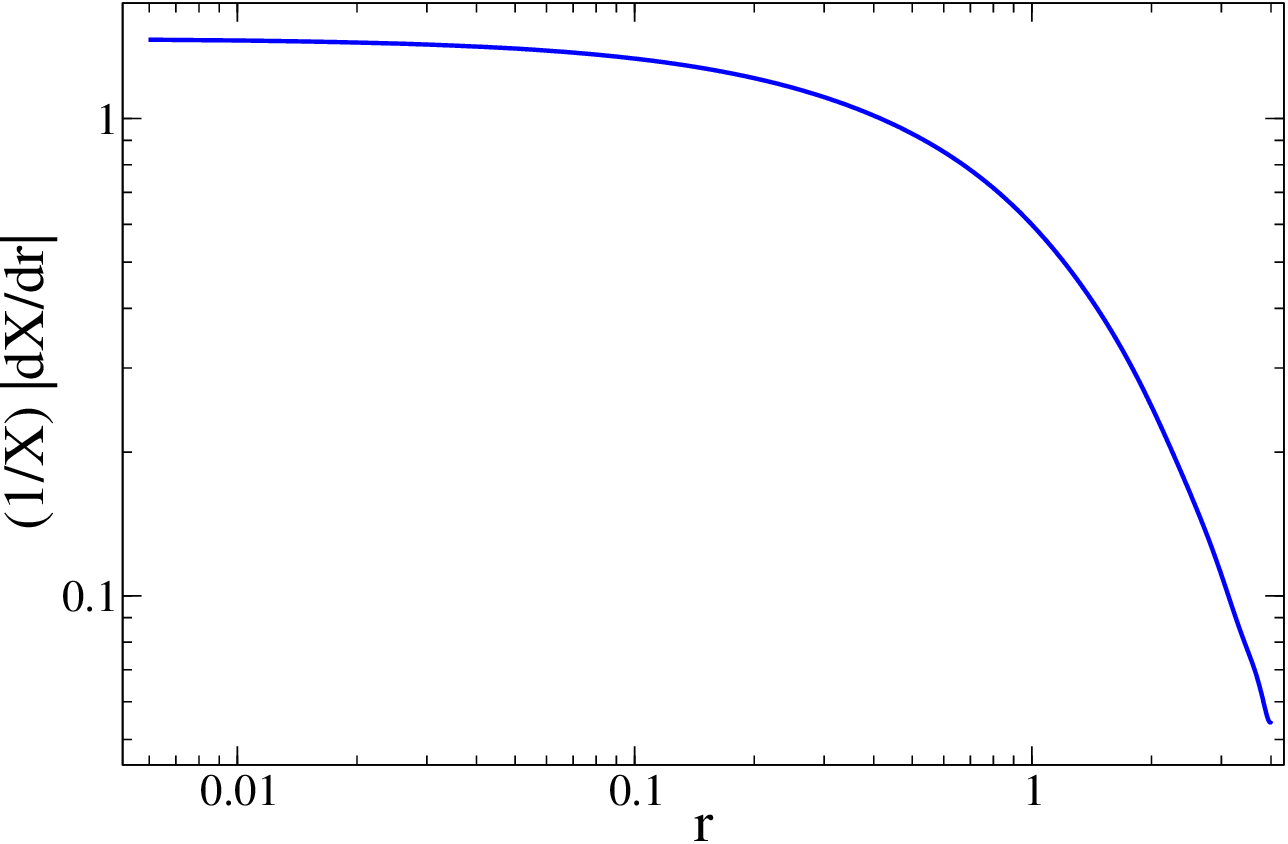}}
\caption{Log-log plot of $X'/X$ versus $r$, whose slope indicates
 that $X'/X$ falls faster than $1/r^2$ (but not faster than $1/r^3$)
 at large $r$. }
\label{wavexp}
\end{minipage}
\end{figure}

In principle, the analysis is complicated by the mixing between the
radion $X$ and the scalar fluctuations $\phi_1$.  However it was
shown by numerical solution of the equations in ref.\
\cite{Burgess:2001bn} that this is a negligible effect, and so 
we can neglect the terms proportional to $\phi_1$ and $\phi_1'$ 
in eq.\ (\ref{gwbc}).  Again, we can solve for the radion mass
in the limit of large warping, $r = P\gg 1$, 
\begin{eqnarray}
m^2_r&=&\left(\frac{5(\alpha-5)}{\ell^2r^3}-2(\alpha+1)\tilde{H}^2
+\frac{15(\alpha-5)}{8\ell^2r^2}\frac{X'}{X}-\frac{3}{4}(\alpha-1)
\tilde{H}^2\left(\frac{rX^\prime}{X}\right)
\,\right.\nonumber\\
&&-\left.\frac{k_6^2\phi_0^{\prime2}}{4\ell^2}
\left(\frac{rX'}{X}\right)-\frac{k_6^2\phi_0^{'2}\tilde{H}^2}{4r}
\left(\frac{X'}{X}\right)\,\right)\bigg|_{r=P}\label{massws}
\end{eqnarray}
The large-$r$ behavior of the radion wave function $X$ is altered by
the GW field; numerically we obtain a solution which is shown in
figure \ref{wavex}.  One observes that $X'/X$ is negative, and that
$|X'/X|$ falls faster than $1/r^2$, but slower than $1/r^3$  (figure
\ref{wavexp}). The sign and magnitude of $X'/X$ implies that the term
proportional  to $\phi_0'^2$ in (\ref{massws}) is positive and that
it is the dominant contribution which makes $m_r^2>0$ in the absence
of Hubble expansion. However, since the order of  $X'/X$ is always
smaller than $r^{-2}$ the second term in (\ref{massws}) dominates
when $\tilde{H}$ becomes sufficiently large, and the radion becomes
unstable.  

We can compare the results here   to our earlier observations
concerning  the Friedmann relation in the stabilized system, figure
\ref{fws}.  There it is seen that the departure of $H^2$ from linear
dependence starts at an $\alpha$-independent value of $H\sim
100/\ell$.  However in section \ref{FEGW}, we used large values of
the GW field, whereas in the present section we treated it as a
perturbation. To make a meaningful comparison, we have  recomputed
the Friedmann relation in the GW-stabilized system for smaller values
of the GW field, by taking smaller values of the coupling $\lambda$.
Motivated by the results of the unstabilized case we investigate
whether the radion mass  also vanishes at the turning point of the
Friedmann equation in this case.  Taking $\lambda=0.02$ which
corresponds  to $\phi(0)=1.53\times 10^{-15}$ at $H=0$, we solve the
eigenvalue problem eq. (\ref{eqX}-\ref{gwbc}) and find  the radion
mass $m_r^2 = -0.00037$ and the deviation from the standard 
Friedmann equation ${\Delta H^2}/{H^2}=$ 61\% at the turning point 
where the upper and lower branches join, {\it i.e.,}
$(H\ell)^2=0.895$. Although the radion mass is not exactly zero at
the turning point, this could be due to our use of the unperturbed
background solutions  $a_0(r)$ and $f_0(r)$ from solutions without GW
field.  To make  a precise comparison the exact background solutions
with GW field would be needed, which would  require further numerical
investigation.

In summary, the fluctuation analysis for the radion suggests that the
exotic branch of the Friedmann equation is invalidated by the
instability of the radion, even in the model with Goldberger-Wise
stabilization.  Moreover, we have not found examples where the
magnitude of $H^2$ deviates from its GR prediction by more than 60\%
before the onset of the instability.

\section{Application of the Friedmann equation to inflation on the brane}
\label{IFL}

An important application of modified Friedmann equations which has
been used in 5D models is to the study of inflation  on the brane,
beginning with ref.\ \cite{Maartens:1999hf}. There the simplest
chaotic inflation  is studied and the inflaton can be under 4D Planck
mass $M_p$ but still above the 5D  scale $M_5$.  In that work, the
high energy regime $H\sim\rho$ was used to obtain novel results, like
the enhancement of number of e-foldings and reduction in the scale of
the inflaton field in chaotic inflation.  Ref.\
\cite{CLL} further showed that inflation can be sustained
for steeper potentials than conventionally.  

We are not able to achieve such dramatic effects in the present
model, because of our observation that the radion becomes unstable
before the deviation from the standard Friedmann equation becomes
very large.  However, we will now show that interesting effects can
nonetheless occur if inflation starts near the point of instability;
namely, large running of the spectral index and breaking of the
consistency relation between the tensor ratio and spectral index.
To demonstrate these effects, we will
 use the modified Friedmann equation obtained
in section \ref{Feq} to study chaotic inflation on the
3-brane with the potential
\be
V=\frac{1}{2}m^2\varphi^2
\label{cotpt}
\ee
We assume that the inflaton field 
on the 3-brane is varying slowly in time and thus its energy density
affects the cosmological expansion in the same way as the excess
3-brane tension in previous sections; it would be surprising if
corrections to this assumption depended in a discontinuous way on the
value of the equation of state $w=p/\rho$.  
To achieve stability of the
extra dimensions, we will consider the model with 
$\alpha=6$ and focus on its low-energy
branch, which we have shown to be stable.

\begin{figure}[h]
\begin{minipage}{8cm}
\vspace{0.95cm}
\centerline{
\includegraphics[width=8cm]{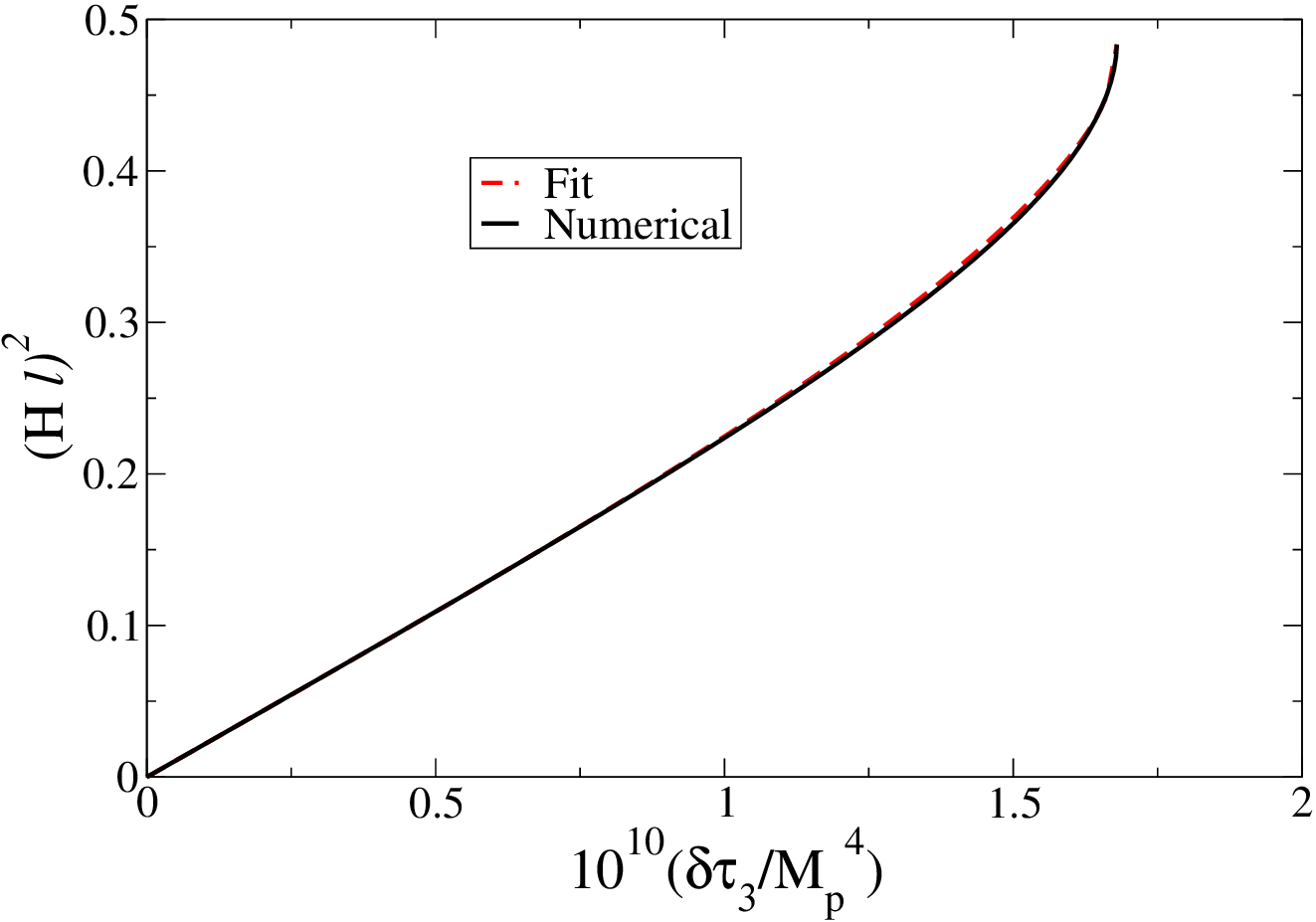}}
\caption{Exact numerical result (solid line) and analytic fit (eq.\
\ref{Ffit}, dashed line) to the Friedmann equation.}
\label{fit}
\end{minipage}
\hspace{2mm}
\begin{minipage}{8cm}
\vspace{0.9cm}
\centerline{
\includegraphics[width=8cm]{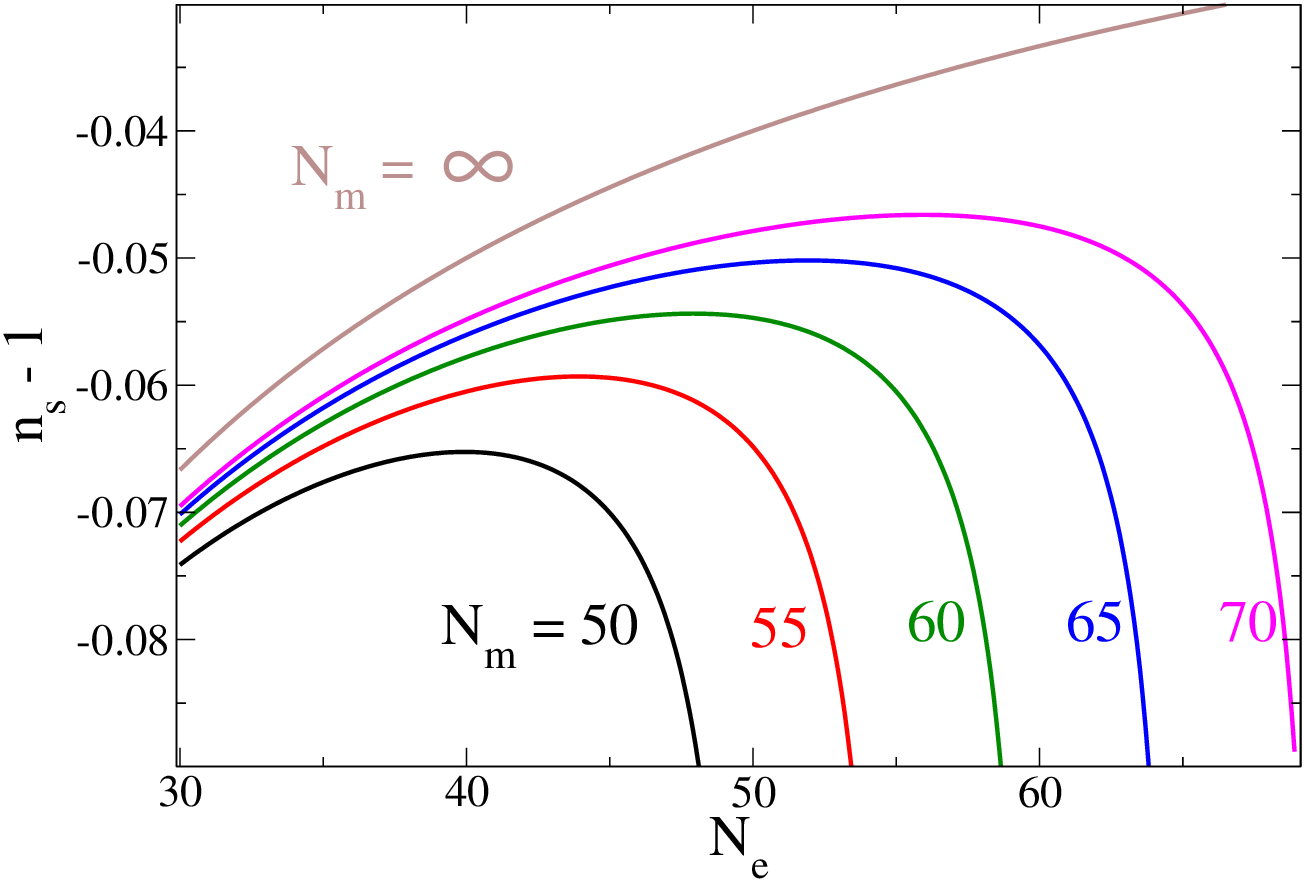}}
\caption{$n_s-1$ versus $N$ from eq.\ (\ref{dns}) for different values of $N_m$, defined in
eq.\ (\ref{Nm}).  The case $N_m=\infty$ is standard chaotic inflation.}
\label{index}
\end{minipage}
\end{figure}

We will denote the modified Friedmann equation by
\begin{eqnarray}
H^2=H_0^2\mathcal{F}(\rho) 
\label{Feqeq}
\end{eqnarray}
$H_0^2={\rho}/{3M_p^2}$  is the standard Friedmann equation and
$\mathcal{F}(\rho)$ quantifies the deviation of $H^2$ from the
standard expression.  Notice that $\mathcal{F}(0)\equiv 1$.  For
the braneworld model we are considering, with $\alpha=6$, we have
found an analytic form for $\mathcal{F}$ which provides
an extremely good fit to the numerical results for the stable
(low-energy) branch:
\begin{eqnarray}
\mathcal{F}(V)=\left(1+f_m-f_m\left(1-\left(\frac{V}{V_m}
\right)^p\right)^{1/p}\right)
\label{Ffit}
\end{eqnarray}
Here $V_m$ is the maximum energy density as shown in figure
\ref{fig13}.  
The fit and the exact numerical result for $H^2$ is plotted in figure
\ref{fit} for a particular set of braneworld parameters,
$T_4=0.00587$, $\ell=1500$, and $\tau_4=4\times10^{-11}$, in units of
$M_6$, which we shall also adopt to illustrate our findings in this
section.  The 4D Planck mass is given by $M_p^2 = 53.8$ in $M_6=1$ units for this
background.   We then find that $\mathcal{F}$ is fit using the values
$f_m=0.33$, $p=2.6$ and $V_m = 4\times 10^{-9} M_p^4$.  We have chosen
this value in order to make the tensor-to-scalar ratio close to the
experimental upper limit, since one of our goals is to show that
deviations in the consistency condition could be an observable
signature of the braneworld model. 

On the other hand, the maximum value of $\mathcal{F}$ is
$1+f_m=1.33$ for these parameters; thus we do not get the ``steep
inflation'' effect discussed by ref.\ \cite{CLL}.
However, this conclusion might change in the scenario where
stabilization is achieved using the Goldberger-Wise mechanism. In the
last section we showed that with a small GW field, $\mathcal{F}\sim
2.5$ at the turning point which is also the dividing point between
the stable and unstable  branches. With a greater GW field the
maximum value of  $\mathcal{F}$ will be even greater and thus it may
be possible to achieve steep inflation.  However in this section we
will focus on the new effects which are unique to our 6D braneworld
model.

\subsection{Spectral index and consistency condition}

We now consider how the modification $\mathcal{F}$ affects inflation.
For our purposes, the main effect is in the relation between the
power spectrum of tensor or scalar perturbations and the inflationary
potential:
\bea
P_{t}&=&\frac{2H^2}{\pi^2 M_p^2} = \mathcal{F}P_{t,0}\nonumber\\
P_{s}&=& {H^4\over 4\pi^2 M_p^2\dot\varphi^2} \cong 
{9H^6\over 4\pi^2 V'^2} = {H^2\mathcal{F}^2 \over 8\pi^2\epsilon M_p^2} =  \mathcal{F}^3P_{s,0}
\label{pspp}
\eea
where $P_{s,0}$ and $P_{t,0}$ are the conventionally defined 
power spectra when expressed in terms of $V$, and $\epsilon$ 
is the usual slow-roll parameter,
\be
\epsilon=\frac12 M_p^2 \left(V'\over V\right)^2
	= {2 M_p^2\over\varphi^2}
\label{eps}
\ee
The factors of $\mathcal{F}$ 
introduce additional
scale dependence in the power spectra beyond that contained in 
$P_{s,0}$ and $P_{t,0}$.  This shows up in modifications to the
spectral indices, 
\begin{eqnarray}
\label{dns}
n_{s}-1&\equiv& \frac{d\ln P_{s}}{d\ln k}={1\over
{\mathcal{F}}}
\left(2\eta - 6\epsilon \left(1+\frac{d\ln \mathcal{F}}{d\ln V}\right)
\right)\\
n_{t}&\equiv& \frac{d\ln P_{t}}{d\ln k}=-{2\epsilon\over{\mathcal{F}}}
\left(1+\frac{d\ln \mathcal{F}}{d\ln V}\right)\label{dnt}\\
r_t&\equiv& \frac{P_{t}}{P_{s}}=\frac{16\epsilon}{\mathcal{F}^2}
\label{tscr}
\end{eqnarray}
These can be derived using the horizon-crossing condition
$k/H=a=e^N$ and the relation between the number of e-foldings $N$
and $\varphi$,
\begin{eqnarray}
N(\varphi)=-\int^{\varphi}_{\varphi_i}
 \frac{\mathcal{F}V}{M_p^2V'}d\varphi
= \frac{\langle\mathcal{F}\rangle}{4M_p^2}\left(\varphi_i^2 -\varphi^2
\right)
\label{efd}
\end{eqnarray}
where the exact value of $\langle\mathcal{F}\rangle$ is not crucial
for our purposes, since $\mathcal{F}$ only varies between 1 and 1.33;
however we can estimate it as 
\be
\langle\mathcal{F}\rangle \cong {1\over
V_m}\int_0^{V_m} \mathcal{F}\, dV \cong 1.05
\ee
Henceforth we set $\langle\mathcal{F}\rangle$ to unity.
These relations imply
\be
{d\ln k} = {dN} + \frac12 {d\ln H^2} = 
\left(-{\mathcal{F}V\over M_p^2 V'} + {V'\over 2V}
\left(1 + {d\ln\mathcal{F}\over d\ln V}\right)\right) {d\varphi}
\cong -{\mathcal{F}V\over M_p^2V'^2}\,dV
\label{dlnk}
\ee
In the slow-roll approximation, this is dominated by 
the first term, $d\ln k \cong -(\mathcal{F}V/M_p^2V'^2)dV$, which
allows us to carry out the differentiations leading to 
eqs.\ (\ref{dns}-\ref{tscr}). 

Before investigating the new effects due to $\mathcal{F}$, we will
impose observational constraints on the inflation model. For 
chaotic inflation, these are conveniently expressed in terms of 
the total number of e-foldings, 
\be 
N_e = N_e(\varphi_f) = 
\frac{1}{4}\left({\varphi_i^2\over M_p^2} -
2\right) \cong \frac{1}{4}\left({\varphi_i^2\over M_p^2}\right)
\label{Ne}
\ee
where we used $\varphi_f^2 = 2M_p^2$ by assuming inflation ends
when $\epsilon=1$.  Therefore the initial value of the inflaton is
given by $\varphi_i^2/M_p^2 = 4N_e$.
The COBE normalization on $P_s$ gives \cite{Peiris}
\be
	{V\over\epsilon M_p^4} = {m^2\varphi_i^4\over 4M_p^6} \cong \tilde A \equiv 7.1\times 10^{-7}
	A(k_0)
\label{COBE}
\ee
where $A(k_0) \cong 0.7-1.0$ is  amplitude at the pivot scale
$k_0$, which we assume corresponds to the field value $\varphi_i$. 
Eqs.\ (\ref{Ne}-\ref{COBE}) imply
\be
	{m^2\over M_p^2} \cong {\tilde A \over 4 N_e^2}
\label{m2eq}
\ee
Furtherfore eqs.\ (\ref{eps}, \ref{tscr}, \ref{Ne}) show that
the tensor-to-scalar ratio is given by 
\be
	r_t = {8\over \mathcal{F}^2 N_e}
\label{rtrt}
\ee

One consequence of the new dependence on $\mathcal{F}$ is that
the spectral index depends differently on the 
number of e-foldings than in standard chaotic inflation, where
$n_s -1 = -2/N_e$.  From eq.\ (\ref{dns}), we obtain the modified
dependence which is graphed in figure \ref{index}, for different
values of $V_m$, which we parameterize as
\be
	V_m = \frac12 m^2\varphi_m^2 = 2 m^2 M_p^2 N_m
\label{Nm}
\ee
Here the parameter $N_m$ is the maximum number of e-foldings due
to the limitation $V<V_m$, and we used the relation  
$\varphi/M_p^2 = 4N$ between the field value and the number of 
e-foldings.  Interestingly, there is an upper limit
on $n_s$ depending on $N_m$ which is significantly smaller than 
$1-2/N_m$.

\begin{figure}[h]
\begin{minipage}{8cm}
\vspace{0.5cm}
\centerline{
\includegraphics[width=8cm]{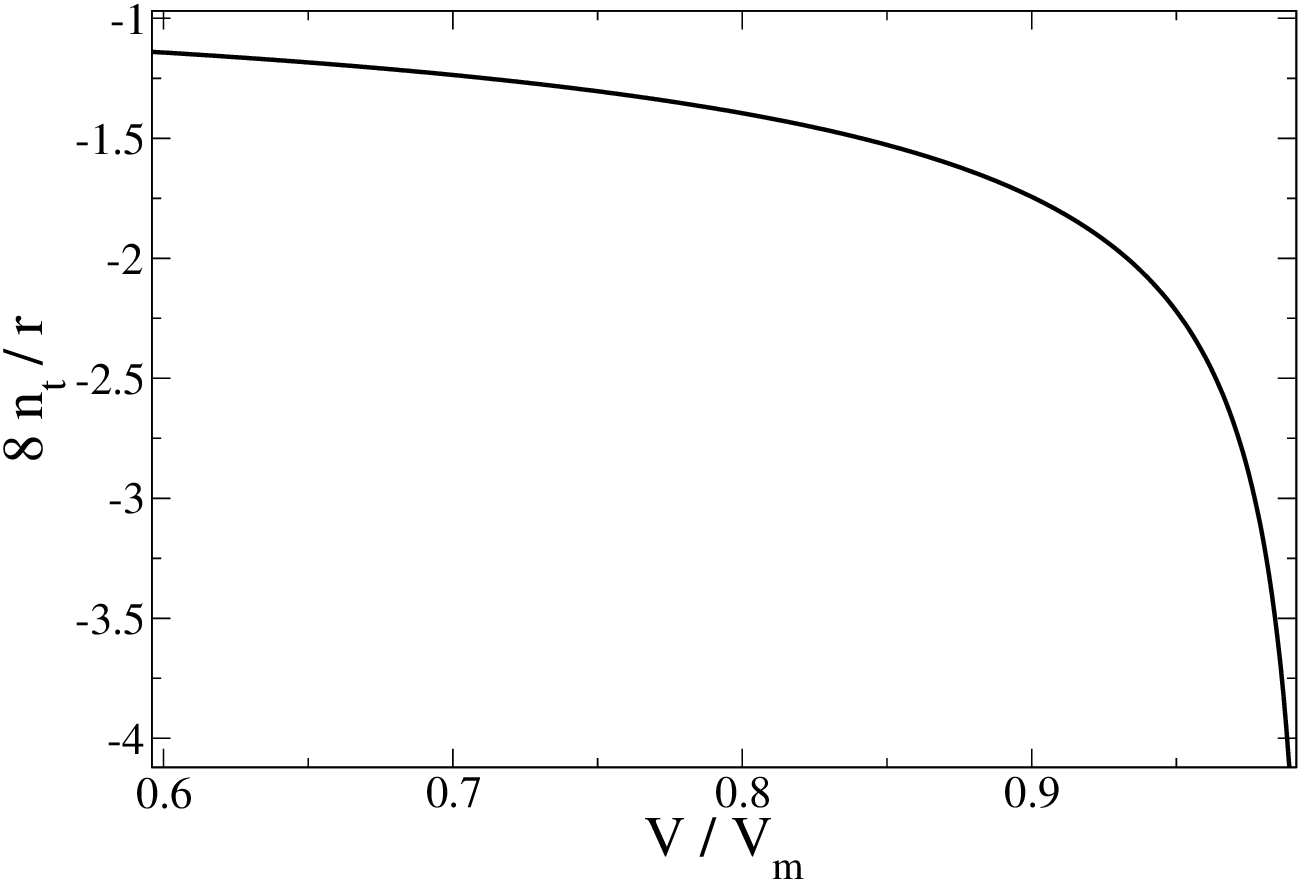}}
\caption{$8 n_t/r_t$ versus $V/V_m$, showing strong deviations from the
standard predicted value of $-1$ as $V\to V_m$.}
\label{ntr2}
\end{minipage}
\hspace{2mm}
\begin{minipage}{8cm}
\vspace{0.95cm}
\centerline{
\includegraphics[width=8cm]{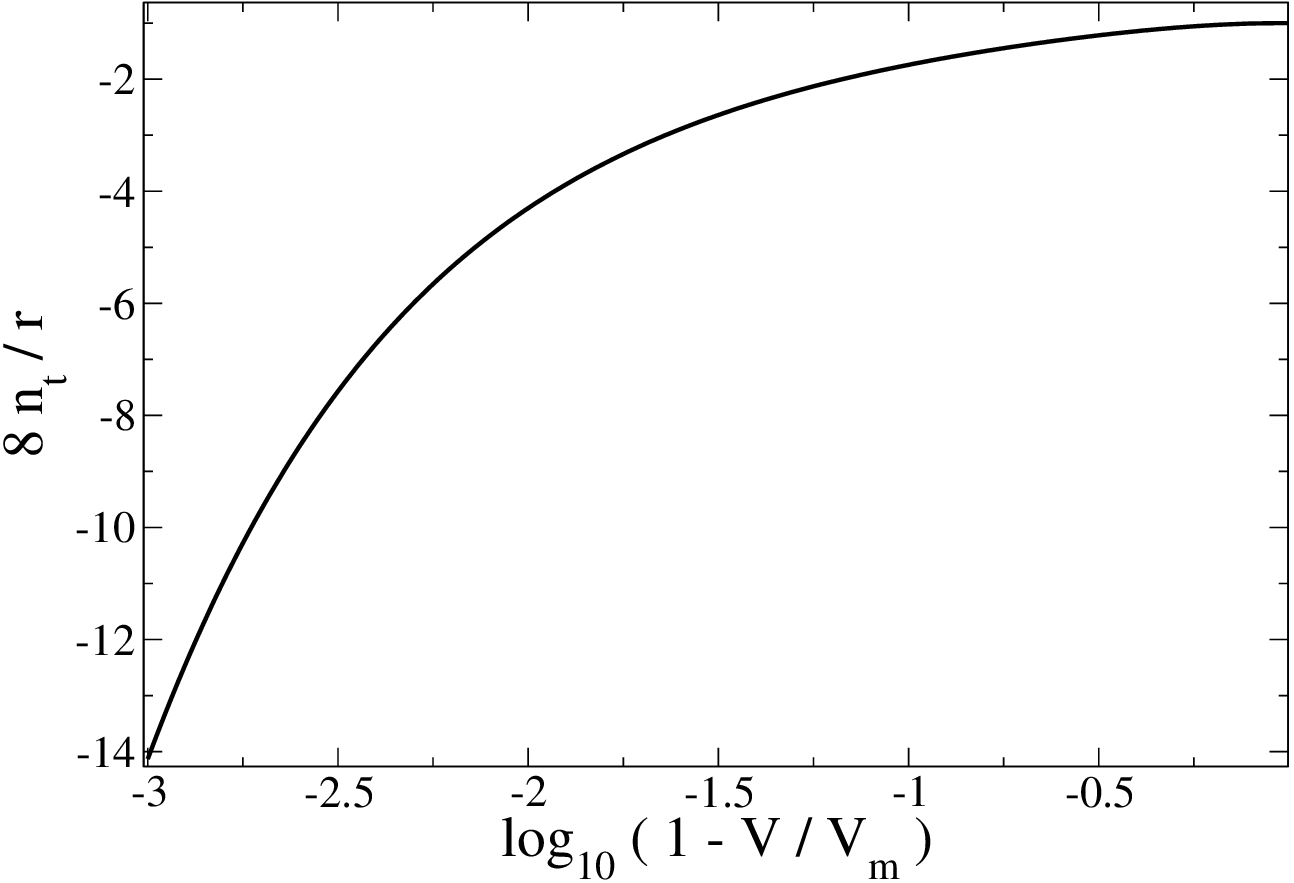}}
\caption{$8 n_t/r_t$ versus $\log_{10}(1 - V/V_m)$, further illustrating
the deviation from the consistency condition when $V$ is very close
to $V_m$.}
\label{ntr3}
\end{minipage}
\end{figure}

A further novel feature is that the 
consistency condition between the tensor spectral index and the
 tensor-to-scalar ratio is different from the conventional prediction,
\be
	{n_t\over r_t} 
=-\frac18{\mathcal{F}\left(1+\frac{d\ln \mathcal{F}}{d\ln V}\right)}
\ee
Using the expression (\ref{Ffit}) for $\mathcal{F}$, we obtain
sizeable deviations of $n_t/r$ from the standard prediction ($-1/8$)
when the inflationary potential $V$ starts out being moderately 
tuned to the maximum value $V_m$ allowed by the brane model.  The
deviations as a function of $V$ are shown in figures \ref{ntr2} and 
\ref{ntr3}.

\begin{figure}[h]
\begin{minipage}{8cm}
\vspace{0.5cm}
\centerline{
\includegraphics[width=8cm]{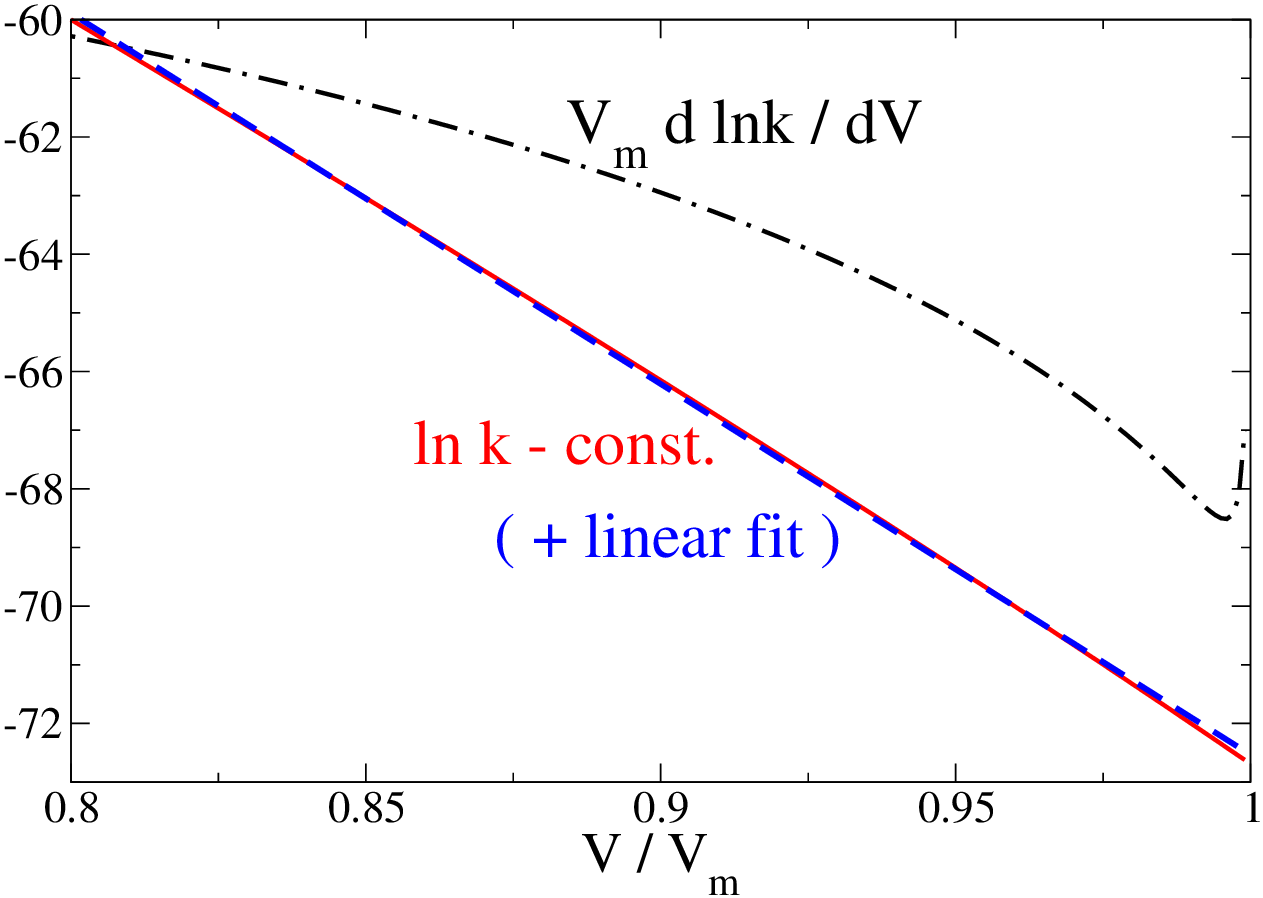}}
\caption{$V_m d\ln k/dV$ versus $V/V_m$, and the integrated result
for $\ln k$ (plus an arbitrary constant, to put it on the same
graph) versus $V/V_m$.  Solid curve is $\ln k$, dashed is the linear
fit.}
\label{dlnkfig}
\end{minipage}
\hspace{2mm}
\begin{minipage}{8cm}
\vspace{0.35cm}
\centerline{
\includegraphics[width=8.2cm]{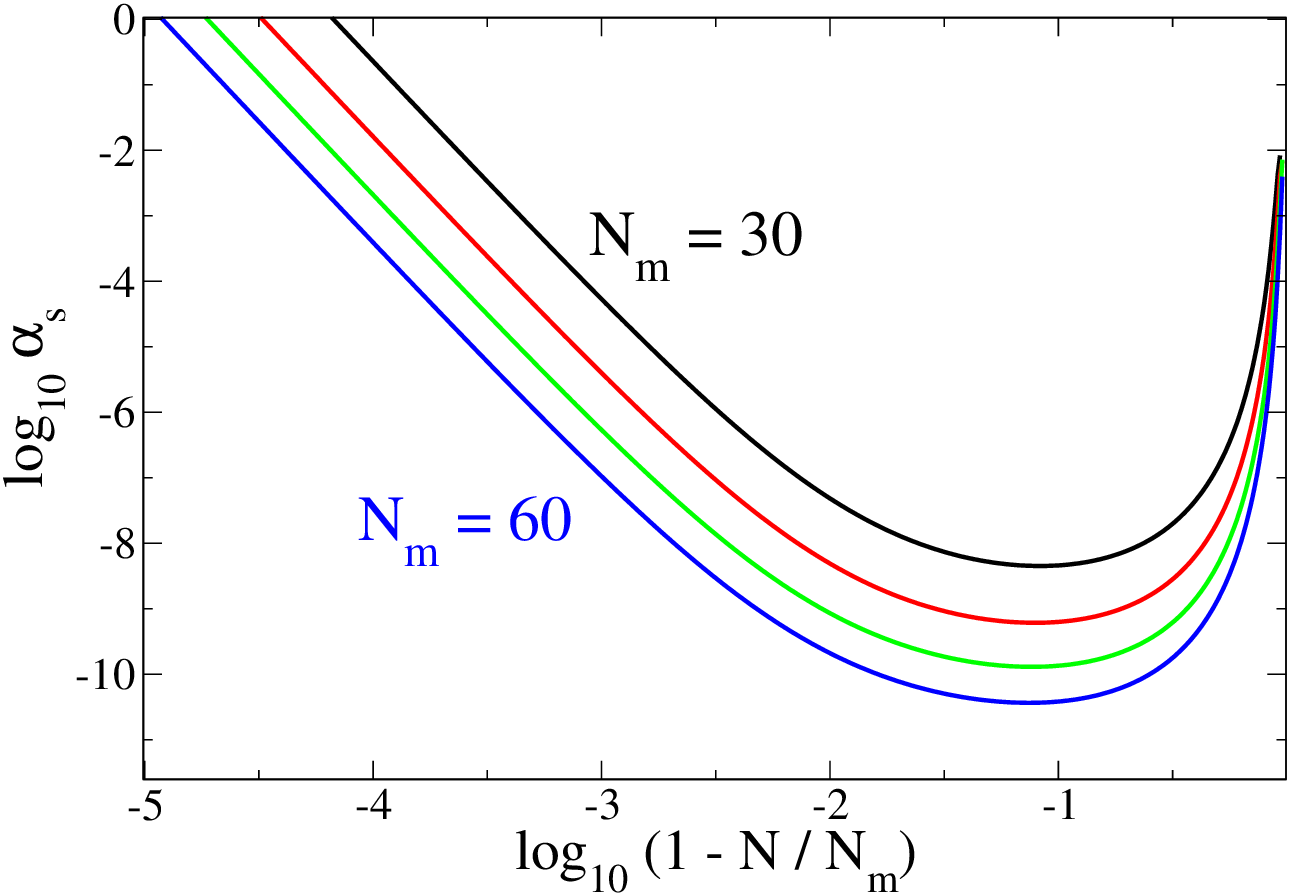}}
\caption{$\log_{10}\alpha_s$ versus $\log_{10}(1-N/N_m)$ for
$N_m = 30,40,50,60$ (top to bottom curves) in the $m^2\varphi^2$
chaotic inflation model.}
\label{running}
\end{minipage}
\end{figure}

The dependence of $n_t/r_t$ on wave number in the CMB spectrum is the 
more observationally relevant issue.  We can analytically elucidate
this dependence using the fact that $\mathcal{F}$ itself does not
deviate greatly during inflation, even though its derivative is
large.  The approximation in eq.\ (\ref{dlnk}) gives $d\ln k/dV
\cong  -\mathcal{F}/(2m^2 M_p^2)$, so that $\ln k$ is approximately
linear in the potential itself.  To demonstrate this, we plot the
exact dependence of $V_m d\ln k/dV$,
\be
	V_n {d\ln k\over dV} = -\mathcal{F}\left(N_e+\frac12\right) + 
	{V_m\over 2V}\left(1 + \frac{d\ln \mathcal{F}}{d\ln V}\right)
\ee
 in figure \ref{dlnkfig}, and
the integrated result for $\ln k$ as a function of $V$, for $N_e=55$
(for other values of $N_e$ the result looks very similar).   The
figure shows that $\ln k$ is to a very good approximation proportional
to $V$.  Therefore the wave-number dependence of $n_t/r$ can be
inferred from figures \ref{ntr2} and \ref{ntr3}.   

\subsection{Running of spectral index}

Another possible spectral feature which can be enhanced by the 
braneworld modification is running of the spectral index, $\alpha_s = 
dn_s/d\ln k$.  From eqs.\ (\ref{dlnk}) and (\ref{dns}) we find
\begin{eqnarray}
\alpha_s=\frac{1}{\mathcal{F}^2}\left(16\epsilon\eta-24\epsilon^2
-2\frac{M_p^4V^\prime V^{\prime\prime\prime}}{V^2}
+16\epsilon\eta\frac{d\ln \mathcal{F}}{d\ln V}
-36\epsilon^2\frac{d\ln \mathcal{F}}{d\ln V}
+12\epsilon^2\frac{d^2\ln \mathcal{F}}{d\ln V^2}-
12\epsilon^2\left(\frac{d\ln \mathcal{F}}{d\ln V}\right)^2\right)
\label{alphas}
\end{eqnarray}
This reduces to the standard expression when  $\mathcal{F}=1$. 
But when the potential is close to its maximum value, $\alpha_s$
is enhanced by the large values of the derivatives of $\mathcal{F}$.
This is potentially interesting because in standard 
inflation, $\alpha_s$ is of higher order in the slow roll parameters,
and thus negligibly small, whereas observationally there is marginal
evidence for running at the level of $\alpha_s\sim -0.1$.  

However, in the chaotic inflation model, we find that $\alpha_s$ is
still negligible unless $V$ is extremely fine-tuned to be close to 
the maximum value $V_m$ (equivalently, $N$ must be very close to
$N_m$).  As illustrated in figure \ref{running},  the required tuning
is worse than 1 part in $10^4$ for $N>40$, which includes the
preferred values of $N$ for the chaotic inflation model.  Furthermore,
eq.\ (\ref{alphas}) predicts that $\alpha_s>0$ due to the enhancement, 
whereas the marginal experimental indications are for negative values
of the running.

\section{CONCLUSION}
\label{CON}

In this paper we have studied the warped codimension-two braneworld
model which most closely resembles the original codimension-one
warped scenario of Randall and Sundrum, with the goal of
understanding how cosmological expansion (the Friedmann equation) is
altered due to the extra dimensions.  The bulk is approximately 
AdS$_6$ with a 3-brane at the bottom of the throat and a 4-brane at
the top; in addition to tension, the 4-brane has an extra
nontensional source whose pressure in the angular direction varies
with the brane circumference $\sim L_4$ as $L_4^{-\alpha}$.  This
extra source is needed in order to find solutions with localized 
gravity when no stabilization mechanism is included, and it is known
\cite{Burgess:2001bn} that the radion is stable when $\alpha> 5$.  It
generally  plays a less important role when stabilization is induced,
for example, using the Goldberger-Wise (GW) mechanism.  

We determined the modified Friedmann equation both analytically,
in a perturbative expansion in powers of the energy density on
the brane, and numerically.  The perturbative result, eq.\ (\ref{FRW35}),
has the curious feature that $H^2 \sim \rho + O(\rho^3)$, {\it i.e.,}
the $O(\rho^2)$ correction vanishes.  For larger values of $\rho$, 
the numerical solution is necessary, as described in section
\ref{NFE}.  We checked that the two kinds of solutions agree with each
other in the small-$\rho$ region where both are valid.

In our study of the modified Friedmann equation for this model, we
have corrected a mistaken claim of ref.\ \cite{CDGV}, which did not
recover the standard GR result at low energies.   In fact we find
standard cosmology at low energy even for the unstabilized models. 
At high energies, we find exotic features, in which the Hubble rate 
$H(\rho)$ turns over and joins with another branch of the function at
some critical value of $\rho$ and $H$,  making $H(\rho)$
double-valued---see figures \ref{alpha} and \ref{fws}.\footnote{The
same phenomenon occurs in the similar model of ref.\ \cite{Send}, but
it is not commented upon there.}  This behavior even persists in the
model with GW stabilization, but we have found that the exotic branch
is nevertheless unphysical, because the model becomes destabilized by
the Hubble expansion  at the turning point.  We discovered this by
doing a small fluctuation analysis around the de Sitter solutions in
section \ref{STB}, in which the eigenvalue problem for the radion
mass $m_r^2$ was solved both using analytical approximations and
exact numerical methods.  This result  is reassuring, since otherwise
we would be left with the puzzle of why the cosmological background
is not uniquely determined by the sources of stress energy in the
Einstein equation.  

Nevertheless, we have found that interesting deviations from the 
standard Friedmann equation occur near the threshold for radion
instability: $dH^2/d\rho$ diverges at this point, even though $H$
itself is finite.  In fact for the specific example we considered, 
$H$ only differs from its standard value by 30\% at this point, but
the divergence in $dH^2/d\rho$ can have observable effects if
inflation happened to start near the maximum value of the potential
allowed by radion stability.  We demonstrated such effects in the
simplest model of chaotic inflation on the 3-brane. Figure
\ref{index} shows that the spectral index is lower than in standard
chaotic inflation; for instance with $N_e=55$ e-foldings, $n_s=0.964$
for standard chaotic inflation, whereas it is $0.950$ for the
braneworld model with $N_m = 65$ as the maximum number of e-foldings
(hence the maximum value of the potential is $65/55$ times its value
at horizon crossing).  Thus without much fine tuning, the braneworld
model can have a potentially measurable impact on the spectral
index.  

We then explored the breaking of the  consistency condition between
the tensor-to-scalar ratio $r_t$  and the tensor spectral index
$n_t$, which may be observable if $r_t$ is close to its experimental
upper limit.  Figure \ref{ntr2} shows that $8n_t/r_t$ differs
by 50\% from the standard prediction of $-1$ even when $V$ is only
within 20\% of its maximum value, so again interesting signatures
can occur without excessive fine tuning of the model parameters.

It would be interesting to try to extend our results to a string
theoretic realization of inflation in a warped throat, namely the
Klebanov-Strassler background \cite{KS} with stabilization by fluxes
\cite{GKP}.  Although we have used GW rather than flux stabilization
in our model, ref.\ \cite{Brummer:2005sh} has argued that an
effective field theory description of flux stabilization might
coincide with the GW picture.  So far, efforts in string theoretic
inflation with warped throats have focused on brane-antibrane
\cite{KKLMMT} or DBI \cite{DBI} inflation.  However the effects we
have discussed should also be applicable in string-based
constructions with conventional inflation confined to a brane.  The
phenomenon of radion destabilization at sufficiently high Hubble rate
should also occur in string compactifications, since the
decompactified background is always a solution, and the radion, being
conformally coupled, gets a mass correction of order $H^2$ in the
early universe.  The $H^2\phi^2$ term in the radion effective action
thus drives $\phi$ away from its nontrivial minimum toward the
decompactified $\phi=0$ vacuum for sufficiently large $H$, and it is
likely that deviations from the usual Friedmann equation will be
large near this point.  Therefore it is possible that stringy effects
due to the extra dimensions may be manifested during inflation even
if the inflaton is confined to the standard model brane.  

\acknowledgements
We wish to thank Guy Moore, Jiro Soda and Misao Sasaki for fruitful discussions.
SK is supported by a grant for research abroad by the JSPS (Japan).  JC and
FC are supported by NSERC (Canada) and FQRNT (Qu\'ebec).
\appendix

\section{Detailed derivation of Modified Friedman Equation}
\label{DD}
We need to determine the higher order terms in
the expansion of $\bar{P}$ in powers of ${\cal H}$.
Again differentiating eq.\ (\ref{JC2:4-brane:1}), we obtain
\begin{eqnarray}
\frac{d^2\bar{P}}{d{\cal H}^2}&=& 2\bar{P}^4\bigg[ \left( -160+32\,{\alpha}^{3}-288\,\alpha-96\,{\alpha}^{2} 
\right) {\bar{P}}^{18}+ \left(
-800+288\,{\alpha}^{2}-480\,\alpha-32\,{\alpha}^{3}
 \right) {\bar{P}}^{15}\nonumber\\ &+& \left( 1224\,\alpha+48\,{\alpha}^{2}+264\,{\alpha}^
{3} \right) {\bar{P}}^{13}+ \left( -504\,{\alpha}^{2}-24\,{\alpha}^{3}+600+
3000\,\alpha \right) {\bar{P}}^{10}+ \left( -633\,{\alpha}^{2}-549\,{\alpha}
^{3}-159\,\alpha-195 \right) {\bar{P}}^{8}\nonumber\\&+& \left( -2160\,\alpha-1056\,{
\alpha}^{2}+144\,{\alpha}^{3} \right) {\bar{P}}^{5}+ \left( -402\,\alpha+230
+306\,{\alpha}^{2}+378\,{\alpha}^{3} \right) {\bar{P}}^{3}-50+522\,{\alpha}^
{2}+162\,{\alpha}^{3}+390\,\alpha\bigg]\nonumber\\
&& \bigg/ \left[ 25\, \left( 5+2\,{\bar{P}}^{5}\alpha-10\,{\bar{P}}^{5}+3\,\alpha \right) ^{3}\right]
\end{eqnarray}
Substituting $P = \bar{P} +  \frac{d\bar{P}}{d{\cal H}}{\cal H} +   
\frac12\frac{d^2\bar{P}}{d{\cal H}^2}{\cal H}^2 $ into the r.h.s.\ of (\ref{3bt}),
 as advocated in the text, we get the surprising result
\begin{eqnarray}
H^2= \frac{8\pi \bar{G}}{3} \delta\tau_3  +  O(\delta\tau_3^3) \ .
\label{invFRW2}
\end{eqnarray}
Namely,  the second order correction vanishes.  

To find the leading correction, we need to go the higher order.  
Continuing the procedure, {\it i.e.}, 
computing ${d^3\bar{P}}/{d{\cal H}^3}$ in (\ref{Pexp}),
using eq.\ (\ref{JC2:4-brane:1}), we find an unwieldy result 
\begin{eqnarray}
\frac{d^3\bar{P}}{d{\cal H}^3}&=& -6\bar{P}^4\bigg[\left(512\,\alpha^5-6656\,\alpha^4+23552\,\alpha^3-89600\,\alpha+5120\,\alpha^2-64000\right)\bar{P}^{28}\nonumber\\
&&+\left(-58240\,\alpha^3+156800\,\alpha^2-280000+8512\,\alpha^4-56000\,\alpha-448\,\alpha^5\right)\bar{P}^{25}\nonumber\\
&&+\left(-99840\,\alpha^2+106240\,\alpha^3+5760\,\alpha^5+630400\,\alpha+64000-51200\,\alpha^4\right)\bar{P}^{23}\nonumber\\
&&+\left(54000\,\alpha+43200\,\alpha^4+64800\,\alpha^3-32400\,\alpha^5-129600\,\alpha^2\right)\bar{P}^{21}\nonumber\\
&&+\left(12880\,\alpha^4-655200\,\alpha^2+54880\,\alpha^3+490000-1680\,\alpha^5+1246000\,\alpha\right)\bar{P}^{20}\nonumber\\
&&+\left(-194080\,\alpha^3+38160\,\alpha^4-94000+5040\,\alpha^5-570000\,\alpha-495840\,\alpha^2\right)\bar{P}^{18}\nonumber\\
&&+\left(-82350\,\alpha^4-47250\,\alpha-180900\,\alpha^3-33750+76950\,\alpha^5+267300\,\alpha^2\right)\bar{P}^{16}\nonumber\\
&&+\left(-47040\,\alpha^4+385280\,\alpha^3-280000-2016000\,\alpha-336000\,\alpha^2\right)\bar{P}^{15}\nonumber\\
&&+\left(35280\,\alpha^4+170000+611360\,\alpha^2+19600\,\alpha-10800\,\alpha^5+485280\,\alpha^3\right)\bar{P}^{13}\nonumber\\
&&+\left(-145800\,\alpha^2+35100\,\alpha^4-67500\,\alpha+67500+167400\,\alpha^3-56700\,\alpha^5\right)\bar{P}^{11}\nonumber\\
&&+\left(1225000\,\alpha+7560\,\alpha^5+35000+35280\,\alpha^3+1089200\,\alpha^2-98280\,\alpha^4\right)\bar{P}^{10}\nonumber\\
&&+\left(84700\,\alpha-1620\,\alpha^5-111500-134460\,\alpha^4-136440\,\alpha^2-356040\,\alpha^3\right)\bar{P}^8\nonumber\\
&&+\left(4050\,\alpha^4-33750+60750\,\alpha+8100\,\alpha^2+12150\,\alpha^5-51300\,\alpha^3\right)\bar{P}^6\nonumber\\
&&+\left(-18900\,\alpha^4-318500\,\alpha-279720\,\alpha^3+17500+11340\,\alpha^5-558600\,\alpha^2\right)\bar{P}^5\nonumber\\
&&+\left(-9360\,\alpha^2+60048\,\alpha^3+13608\,\alpha^5+23000+56376\,\alpha^4\bar{P}^3-12600\,\alpha\right)\bar{P}^3\nonumber\\
&&+33453\,\alpha^4+5103\,\alpha^5+28875\,\alpha+81270\,\alpha^3+85050\,\alpha^2-4375\bigg]\nonumber\\
&&\bigg/\bigg[125(-10\bar{P}^5+2\,\alpha \bar{P}^5+5+3\,\alpha)^5\bigg]
\label{d3PdH3}
\end{eqnarray}

Substituting $P = \bar{P} +  \frac{d\bar{P}}{d{\cal H}}{\cal H} +   
\frac12\frac{d^2\bar{P}}{d{\cal H}^2}{\cal H}^2 +   
\frac16\frac{d^3\bar{P}}{d{\cal H}^3}{\cal H}^3  $ into the r.h.s.\ of (\ref{3bt}), 
we get
\begin{eqnarray}
\frac{k_6^2}{2\pi}\delta\tau_3=\frac{2}{5}\left(\,1-\frac{k_6^2}{2\pi}\bar\tau_3\,\right)(\bar{P}^3-1)
\left({\cal H}-{\cal H}^3 J(\alpha,{\cal H})+O({\cal H}^4)\right)\nonumber\\
\label{invFRW3}
\end{eqnarray}
where
\begin{eqnarray}
\label{Jeq}
J(\alpha,{\cal H}) &=&
\bigg[ (-1120-1056\,\alpha+96\,\alpha^2+32\,\alpha^3)\bar{P}^{24}+(5088\,\alpha+600
+144\,\alpha^3+312\alpha^2)\bar{P}^{19}\nonumber\\
&&+(3000-24\,\alpha^3+360\,\alpha^2-1800\,\alpha)\bar{P}^{18}+(-8\,\alpha^3+120\,\alpha^2-600\,\alpha+1000)\bar{P}^{15}\nonumber\\
&&+(-459\,\alpha^3-165-3087\,\alpha^2-2433\,\alpha)\bar{P}^{14}+(-108\,\alpha^3-4500+900\,\alpha^2-900\,\alpha)\bar{P}^{13}\nonumber\\
&&+(-36\,\alpha^3-1500+300\,\alpha^2-300\,\alpha)\bar{P}^{10}+(185+1179\,\alpha^2+783\,\alpha^3-99\,\alpha)\bar{P}^9\nonumber\\
&&+(2250\,\alpha-162\,\alpha^3+270\,\alpha^2+2250)\bar{P}^8+(90\,\alpha^2+750-54\,\alpha^3+750\,\alpha)\bar{P}^5\nonumber\\
&&+(-405\,\alpha^2-81\,\alpha^3-375-675\,\alpha)\bar{P}^3-225\,\alpha-125-27\,\alpha^3-135\,\alpha^2\bigg]\nonumber\\
&&\bigg/\bigg[25\left((2\,\alpha - 10)\bar{P}^5+5+3\,\alpha\right)^3(\bar{P}^3-1)
                             \bigg]   \ .
\end{eqnarray}

\newpage
\section{Symbols Used in the Paper}
\label{sec:symbols}
\begin{table*}[!h]
\caption{Definitions of symbols used in this paper (A-M) }
\begin{center}
\begin{tabular}{cll}
\hline
Variable &  Description & Definition or first appearance\\
\hline \hline
$A$ & amplitude of CMB fluctuations & eq.\ (\ref{COBE})\\
$\tilde A$ & COBE normalization parameter & eq.\ (\ref{COBE})\\
$A_0', B_0', C_0'$ & logarithmic derivative of $a_0,b_0,c_0$ &  above
eq.\ (\ref{eqX}) \\
$a,f$ & metric components in $r$ coordinate & eq.\ (\ref{metric1})\\
$a,b,c$ & metric components in stability analysis & eq. (\ref{sfluc})\\
$a_0, b_0, c_0$ & unperturbed solutions for $a,b,c$ & eqs.\
(\ref{a0eq})-(\ref{c0eq})\\
$a_1, b_1, c_1$ & perturbations of $a,b,c$ & eqs.\
(\ref{a0eq})-(\ref{c0eq}) \\
$\alpha$ & eq.\ of state parameter for $\tau_4$ & eq.\
(\ref{stresstensor}) \\
$\alpha_s$ & running of spectral index & eq.\ (\ref{alphas}) \\
$b$ & parameter of bulk potential & eq.\ (\ref{bulkV}) \\
$c$ & dimensionless 4-brane tension parameter & $c=(k_6^2\ell T_4)^2$\\
$\delta{\tau_3}$ & deviation of the 3-brane tension from $\bar{\tau_3}$ & $\delta{\tau_3}=\tau_3-\bar{\tau_3}$ \\
$\Delta H^2$ & deviation of $H^2$ from standard value & eq.\ (\ref{fracdev}) \\
$\epsilon$, $\eta$ & conventional slow-roll parameters &
$\epsilon={M_p^2}{V'^2}/2{V^2}$, $\eta={M_p^2V''}/{V}$ \\
$\tilde g_{\mu\nu}$, $\tilde g_{\theta\theta}$ & induced metric on 4-brane & eq.\
(\ref{stresstensor})\\
$\mathcal{F}$ & modification factor for Friedmann eq. & eq.\ (\ref{Feqeq}) \\
$G$ & 4D Newton's constant  & eq.\ (\ref{graviconstant1})\\
$\bar G$ & 4D Newton's constant in static solution & eq.\ (\ref{MP4})\\
$H$ & proper time Hubble parameter & above eq.\ (\ref{rhoeq})\\
$H_0$ & standard GR value of $H$ & in the last paragrph of \ref{usm} and eq.\ (\ref{Feqeq})\\
$H_t$ & $H$ at turning point in Friedmann eq. & fig.\ \ref{Hct}\\
$H_c$ & critical $H$ for radion stability & fig.\ \ref{Hct}\\
$\tilde{H}$ & rescaled Hubble parameter & $\tilde{H}=H\sqrt{a(\rho)}$\\
${\cal H}$ & dimensionless squared Hubble parameter$\quad$ & ${\cal H}=H^2\ell^2$ \\
$J$ & $\rho^3$ coefficient in modified Friedmann eq. & eq.\ (\ref{FRW35})\\
$k$ & wave number for CMB fluctuations & eq.\ (\ref{dns})\\
$k_0$  & pivot scale for CMB fluctuations & eq.\ (\ref{COBE})\\
$k_6^2$ & 6D gravitational constant & $k_6^2=M_6^{-4}$, eq.\ (\ref{theaction}) \\
$k^2$  & reduced 6D gravitational constant & $k^2={k_6^2}/{2\pi}$ \\
$K$ & conical deficit parameter & eq.\ (\ref{metric1}) \\
$\ell$ & curvature length scale & $\ell^2=-{10}/{\Lambda_6}$\\
$L_4$, $L(P)\ \ $ & (4-brane circumference)$/2\pi$ & eq.\ (\ref{stresstensor}) \\
$\lambda$ & coupling of GW field to 4-brane & eq.\ (\ref{gwcb})\\
$\Lambda_4$ & 4D cosmological constant & $\Lambda_4=3H^2$\\
$\Lambda_6$ & 6D cosmological constant & eq.\ (\ref{theaction})\\
$m$ & inflaton mass & eq.\ (\ref{cotpt})\\
$m_r$ & radion mass & above eq.\ (\ref{pt:mc})\\
$M,L$ & metric components in $\tilde r$ coordinate & eq.\ (\ref{metric2})\\
$M_4$ & 4D Planck mass & eq. (\ref{graviconstant1})\\
$M_5$ & 5D mass scale & eq.\ (\ref{5dfei})\\
$M_6$ & 6D mass scale &  $M_6^{-4}= k_6^2$\\
$M_p$ & 4D Planck mass & $M_p=M_4$\\
$\bar{M_4}$ & 4D Planck mass in static solution& eq. (\ref{MP4}) \\
\hline
\end{tabular}
\end{center}
\end{table*}

\begin{table*}[!h]
\caption{Definitions of symbols used in this paper (N-X) }
\begin{center}
\begin{tabular}{cll}
\hline
Variable &  Description & Definition or first appearance\\
\hline \hline
$N_e$ & number of e-foldings at horizon crossing & eq.\ (\ref{Ne})\\
$N_m$ & maximum number of e-foldings & eq.\ (\ref{Nm})\\
$n_s$  & scalar spectral index  & eq.\ (\ref{dns})\\
$n_t$ & tensor spectral index & eq.\ (\ref{dnt}) \\
$P$ & 4-brane position in units of $\ell$ &eq.\ (\ref{einstein}) and below eq.\
(\ref{xform})\\
$\bar{P}$ & 4-brane position for the static solution & eq.\ (\ref{OP}) and eq.\ (\ref{casimir})\\
$P_s$ & scalar power spectrum & eq.\ (\ref{pspp})\\
$P_t$ & tensor power spectrum & eq.\ (\ref{pspp})\\
$\phi$ & Goldberger-Wise field & in section \ref{FEGW}\\
$\phi_0$ & unperturbed GW field & in section \ref{gwstab}\\
$\phi_1$ & perturbation to GW field  & in section \ref{gwstab}\\
$\varphi$ & inflaton & eq.\ (\ref{cotpt})\\
$\varphi_i$  & inflaton at horizon crossing & eq.\ (\ref{efd})\\
$\varphi_f$  & inflaton at end of inflation & eq.\ (\ref{Ne})\\
$\rho$ & energy density & eq.\ (\ref{Feqeq})\\
$\varrho$ & 3-brane position in units of $\ell$ & eq.\ (\ref{einstein}) and below eq.\
(\ref{xform}) \\
$r, \tilde r$ & radial bulk coordinates & eq.\ (\ref{metric1}) and eq.\ (\ref{metric2})\\
$r_t$ & tensor to scalar ration & eq.\ (\ref{rtrt})\\
$\tau_3$ & 3-brane tension & eq.\ (\ref{theaction})\\
$\bar{\tau_3}$ & 3-brane tension for the static solution & above eq.\ (\ref{OP})\\
$T_4$ & 4-brane tension & below eq.\ (\ref{theaction})\\
$\tau_4$ & nontensional 4-brane energy density & eq.\ (\ref{stresstensor})\\
$u,v$ & logarithmic derivatives of $M,L$ & $u=M'/M$, $v=L'/L$ \\
$V$ & inflaton potential & eq.\ (\ref{cotpt}) \\
$V_m$ & maximum value of inflaton potential & below eq.\ (\ref{Ffit})\\
$V_0$ & 4-brane energy density & eq.\ (\ref{pot1})\\
$V_\theta$ & 4-brane angular pressure  & eq.\ (\ref{pot2})\\
$V_P$ & bulk scalar potential on the 4-brane & $V_P=-\lambda \phi$\\
$\cal{V}$ & bulk scalar potential & $\mathcal{V}=(\frac{b^2}{2}+\frac{5b}{2\ell})\phi^2-\frac{5}{32}b^2\phi^4-\frac{10}{\ell^2}$\\
$W$ & ``superpotential'' for bulk potential & eq.\ (\ref{spww})\\
$X$ & combination of metric perturbations & eq.\ (\ref{cbox})\\
\hline
\end{tabular}
\end{center}
\end{table*}

\end{document}